\documentclass[aps,prl,reprint,superscriptaddress]{revtex4-2}
\usepackage[utf8]{inputenc}

\newcounter{myequation}
\newcounter{myfigure}
\newcounter{mytable}
\makeatletter
\@addtoreset{equation}{myequation}
\@addtoreset{figure}{myfigure}
\@addtoreset{table}{mytable}
\makeatother

\usepackage[hidelinks]{hyperref}

\usepackage{hyperref}
\usepackage{color}
\hypersetup{colorlinks=true}
\usepackage{graphicx}
\usepackage{bm}
\usepackage{amsmath}
\usepackage{amssymb}
\DeclareMathOperator{\diag}{diag}

\usepackage{physics}
\usepackage{siunitx}
\usepackage{chemformula}

\usepackage{txfonts}

\usepackage[textsize=footnotesize]{todonotes}

\begin{document}

	\title{Zero-frequency chiral magnonic edge states protected by non-equilibrium topology}

	\author{Pieter M. Gunnink}
	\email{p.m.gunnink@uu.nl}
	\affiliation{Institute for Theoretical Physics and Center for Extreme Matter and Emergent Phenomena, Utrecht University, Leuvenlaan 4, 3584 CE Utrecht, The Netherlands}
	
	\author{Joren S. Harms}
	\affiliation{Institute for Theoretical Physics and Center for Extreme Matter and Emergent Phenomena, Utrecht University, Leuvenlaan 4, 3584 CE Utrecht, The Netherlands}
	\author{Rembert A. Duine}
	\affiliation{Institute for Theoretical Physics and Center for Extreme Matter and Emergent Phenomena, Utrecht University, Leuvenlaan 4, 3584 CE Utrecht, The Netherlands}
	\affiliation{Department of Applied Physics, Eindhoven University of Technology, P.O. Box 513, 5600 MB Eindhoven, The Netherlands}
	\author{Alexander Mook}
	\affiliation{Institute of Physics, Johannes Gutenberg-University Mainz, Staudingerweg 7, Mainz 55128, Germany}

	\date{\today}
	\begin{abstract}
		Topological bosonic excitations must, in contrast to their fermionic counterparts, appear at finite energies. This is a key challenge for magnons, as it prevents straightforward excitation and detection of topologically-protected magnonic edge states and their use in magnonic devices. 
		In this work, we show that in a non-equilibrium state, in which the magnetization is pointing against the external magnetic field, the topologically-protected chiral edge states in a magnon Chern insulator can be lowered to zero frequency, making them directly accessible by existing experimental techniques. We discuss the spin-orbit torque required to stabilize this non-equilibrium state, and show explicitly using numerical Landau-Lifshitz-Gilbert simulations that the edge states can be excited with a microwave field. Finally, we consider a propagating spin wave spectroscopy experiment, and demonstrate that the edge states can be directly detected.
	\end{abstract}
	\maketitle
	
	\textit{Introduction. }
	Over the past decade, it has become clear that the concepts of topological band theory cannot only be applied to electrons \cite{haldaneModelQuantumHall1988,hasanColloquiumTopologicalInsulators2010}, but also to a whole range of other (quasi)-particles, encompassing photons \cite{haldanePossibleRealizationDirectional2008,ozawaTopologicalPhotonics2019a} and collective bosonic modes in quantum condensed matter systems like phonons \cite{maTopologicalPhasesAcoustic2019}, plasmons \cite{jinTopologicalMagnetoplasmon2016, jinInfraredTopologicalPlasmons2017}, and magnons \cite{mcclartyTopologicalMagnonsReview2022}. Among the latter, topological magnon systems, such as magnon Chern insulators \cite{katsuraTheoryThermalHall2010,vanhoogdalemMagneticTextureinducedThermal2013,shindouTopologicalChiralMagnonic2013,zhangTopologicalMagnonInsulator2013,mookEdgeStatesTopological2014,owerreFirstTheoreticalRealization2016a,kimRealizationHaldaneKaneMeleModel2016, mookInteractionStabilizedTopologicalMagnon2021}, magnon spin Hall insulators \cite{nakataMagnonicTopologicalInsulators2017, mookTakingElectronmagnonDuality2018,kondoMathbbZTopological2019}, magnon Dirac semimetals \cite{franssonMagnonDiracMaterials2016, pershogubaDiracMagnonsHoneycomb2018}, magnon Weyl semimetals \cite{liWeylMagnonsBreathing2016, mookTunableMagnonWeyl2016}, and higher-order topological magnon insulators \cite{liHigherorderTopologicalSolitonic2019, hirosawaMagnonicQuadrupoleTopological2020, mookChiralHingeMagnons2021} are especially of interest because they couple to external magnetic fields providing an exceptional handle for control. Arguably, the most fundamental of these phases is the magnon Chern insulator, which supports chiral edge states that could be used as fault-tolerant spin-wave current splitters and interferometers \cite{shindouTopologicalChiralMagnonic2013, wangTopologicalMagnonicsParadigm2018} and for highly efficient spin transport robust against backscattering at moderate disorder \cite{ruckriegelBulkEdgeSpin2018,wangBosonicBottIndex2020}.
	Multiple magnetic materials have been predicted to be magnon Chern insulators from their bulk band structure obtained by inelastic neutron scattering experiments \cite{chisnellTopologicalMagnonBands2015, chenTopologicalSpinExcitations2018,zhuTopologicalMagnonInsulators2021,weberTopologicalMagnonBand2022}. However, the hallmark chiral edge states have to date not been directly observed. Alternatively, as a direct probe of bulk band topology Raman scattering has been proposed \cite{vinasbostromDirectOpticalProbe2023}.
	
	\begin{figure}[hb!]
		\centering
		\includegraphics[width=\columnwidth]{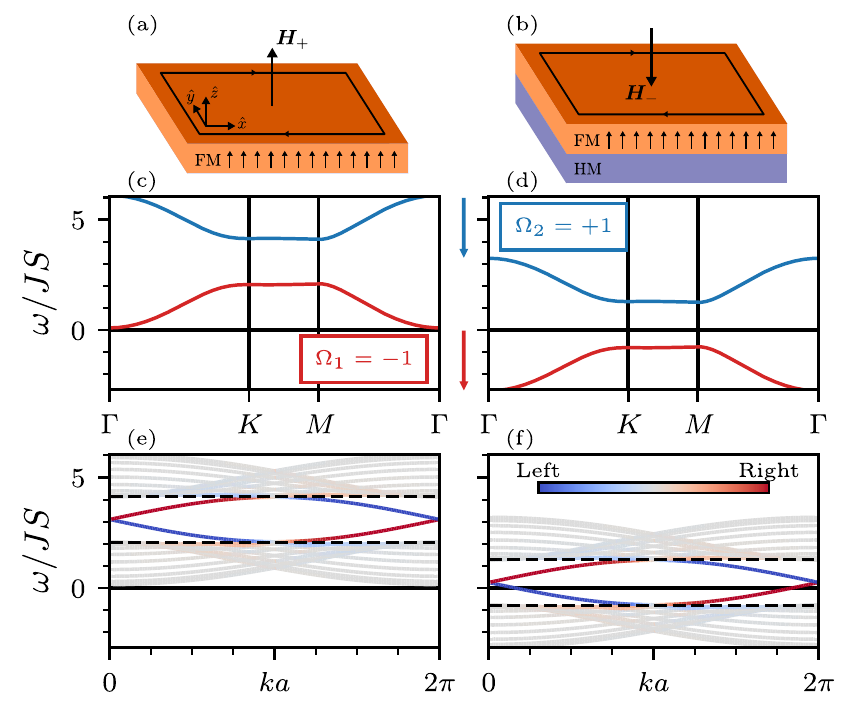}
		\caption{
			Strategy for generating zero-frequency chiral magnonic edge states in a  magnon Chern insulator ferromagnet (FM), comparing the equilibrium (a,c,e) with the non-equilibrium (b,d,f) situation, with the uniform magnetization (see arrows) and magnetic field $\bm{H}_\pm$ aligned parallel and anti-parallel, respectively. In the non-equilibrium case, the magnetization is stabilized by spin-orbit torques originating from the adjacent heavy metal (HM) layer. (c,d) Bulk magnon band structure with indicated Chern numbers, $\Omega_{1,2}$. (e,f) Magnon band structure of an armchair edge ribbon. The colorscale of the eigenfrequencies indicates the edge localization and dashed lines denote the bulk band gap. In equilibrium, $H_+/(JS)=0.1$, the edge states lie at high frequencies, but they are lowered down to zero frequency in non-equilibrium, $H_-/(JS)=-2.75$. 
			\label{fig:main-fig}}
	\end{figure}
	
	This lack of evidence for chiral bosonic edge states is strongly tied to the fundamental difference between fermion topological insulators and their bosonic analogs. 
	Since bosonic collective excitations do not obey a particle number conservation law, their mathematical description relies on the Bogoliubov-de-Gennes formalism, which comes with a doubled particle space. 
	As a result, the topologically-protected edge states have to appear at finite frequencies above the first bulk band \cite{shindouTopologicalChiralMagnonic2013, luMagnonBandTopology2018, xuSquaringFermionThreefold2020}. For magnon Chern insulators, this means the edge states have energies set by the magnetic exchange energy scale, which is typically \si{meV} \cite{menaSpinWaveSpectrumQuantum2014,chenTopologicalSpinExcitations2018}. The corresponding frequency is in the \si{THz}-range, which is beyond the reach of experimental tools, such as Brillouin Light Scattering or microwave excitation and detection.

	In this work we propose a method to lower the topologically-protected chiral edge states in magnon Chern insulators to zero frequency, such that they are easily accessible by microwave techniques. This is achieved by considering magnon excitations on top of a uniform magnetization that is pointing \emph{against} the applied external magnetic field, as opposed to considering excitations on top of a magnetization parallel to the magnetic field, as shown in Fig.~\hyperref[fig:main-fig]{\ref*{fig:main-fig}(a-b)}. In such a \emph{non-equilibrium} setup, the magnon excitations decrease the energy of the system, allowing us to tune the frequency of the edge modes to zero frequency. 
	Since the non-equilibrium state is energetically unstable, it has to be rendered dynamically stable, which is achieved by an appropriate spin-orbit torque. 
	Using numerical Landau-Lifshitz-Gilbert simulations we show that the edge modes can be excited at low frequencies, and are topologically protected against backscattering. Finally, we consider a propagating spin-wave spectroscopy (PSWS) experiment with two antennas, and demonstrate that the edge modes can be directly detected at gigahertz frequencies, even in the presence of disorder.

	\textit{Model. }We consider a two-dimensional magnetic system of localized spins $\bm S_i$ with length $S$ on two sublattices (denoted $\mathcal{A}$ and $\mathcal{B}$), subject to an external magnetic field $\bm H$ with strength $H_0$ and orientated along the $z$-axis, such that $\bm H= H_\pm \hat{\bm{z}}$, where we have introduced $H_\pm\equiv\pm H_0$. After linearizing the spin Hamiltonian $\mathcal H$ in fluctuations around a uniform state we find a two-band spin-wave Hamiltonian. We assume the spin-wave Hamiltonian to realize a magnon Chern insulator, exhibiting topologically non-trivial magnon bands, and topologically-protected chiral edge states whose dispersion run across the bulk band gap. The edge states therefore have a finite energy, which cannot be lower than that of the first bulk band \cite{shindouTopologicalChiralMagnonic2013,luMagnonBandTopology2018, xuSquaringFermionThreefold2020}.
	
	The central thesis of this work is that we can use a non-equilibrium state with the magnetization pointing against the external magnetic field to lower the edge states down to zero frequency. We thus consider the state $\bm S_i=S\hat{\bm{z}}$, whilst $\bm H=H_-\hat{\bm{z}}$. We refer to the case of $H=H_+$ as the equilibrium, and $H=H_-$ as the non-equilibrium. The non-equilibrium state is unstable and will thus relax to the equilibrium state in the presence of dissipation---such as Gilbert damping---with the magnetization parallel to the applied magnetic field. A spin-orbit torque is therefore necessary to render the energetically-unstable situation dynamically stable. Experimentally, this could be accomplished by interfacing the ferromagnetic insulator (FM) with a heavy metal (HM), as indicated in Fig.~\hyperref[fig:main-fig]{\ref*{fig:main-fig}(b)}, such that the spin Hall effect generates a transverse spin current in the HM, injecting spin into the FM \footnote{Because of the specific geometry considered here, where the magnetization is perpendicular to the plane, one would need to make use of the anomalous spin Hall effect in a ferromagnetic heavy metal, such as permalloy \cite{dasSpinInjectionDetection2017}.}. 
	
	The spin dynamics are governed by the semiclassical Landau-Lifshitz-Gilbert (LLG) equation 
	\begin{equation}
		\partial_t \bm S_i = \bm S_i \times \left(-\frac{\partial \mathcal{H}}{\partial \bm S_i} + \frac{\alpha}{S} \partial_t \bm S_i + \frac{J_s}{S}\bm S_i \times  \hat{\bm{z}}\right), \label{eq:LLG}
	\end{equation}
	where $\alpha$ is the Gilbert damping and we allow for the system to be driven by a spin-orbit torque, $J_s$.  We now expand the LLG Eq.~\eqref{eq:LLG} in deviations $m_{\mathcal{A/B},i}^{\pm}=(S_{\mathcal{A/B},i}^x \mp i S_{\mathcal{A/B},i}^y)/\sqrt{2S}$ around the uniform state, $\bm S_{\mathcal{A/B},i} = S\hat{ \bm{z}}$, where $m^\pm_{\mathcal{A/B},i}$ refer to excitations for the equilibrium state, $H=H_+$, and non-equilibrium state, $H=H_-$, on the sublattices $\mathcal{A/B}$. After introducing the Fourier transform of the spin-wave operators, $m^\pm_{\mathcal{A/B},i} = \sqrt{2/N} \sum_{\bm k} e^{i\bm k\cdot \bm R_i}m_{\mathcal{A/B},\bm k}^\pm$,  the LLG Eq.~\eqref{eq:LLG} can  be written as a Bogoliubov-de-Gennes (BdG) like equation in momentum space,
	\begin{equation}
		i(\tau_0 + i\alpha \tau_z)\partial_t \bm\Psi_{\bm k}^\pm = (\tau_z \bm{\mathcal H}_{\bm k}^\pm + i J_s \tau_0 ) \bm \Psi_{\bm k}^\pm,
		\label{eq:bdg}
	\end{equation}
	where $\tau_{\eta}$ are the Pauli matrices in particle-hole space and we have introduced the magnon state vector $\bm \Psi_{\bm k}^\pm = (m^\pm_{\mathcal{A},\bm k}, m^\pm_{\mathcal{B},\bm k,} m_{\mathcal{A},-\bm k}^{\pm*}, m^{\pm *}_{\mathcal{B},-\bm k} )^T$ in particle-hole space.

	We first determine the stability criterion for the non-equilibrium state, which can be found by solving the BdG-like Eq.~(\ref{eq:bdg}) up to zeroth order in $\bm k$ and up to first order in the dissipative terms, $\alpha$ and $J_s$. We then find that $\omega_{0,\pm} = H_\pm - i (\alpha H_\pm-J_s)$. For stability, we require that $\Im[\omega_{0,\pm}]<0$, which in equilibrium, where $H=H_+>0$, means that the system is stable in the absence of spin-orbit torque. In non-equilibrium, where $H=H_-<0$, we require that $J_s \ge \alpha H$ and thus the non-equilibrium state can be rendered dynamically stable with a sufficiently large spin-orbit torque. 
	
	Although our general method is valid for any magnon Chern insulator, we now explicitly consider the well-known magnon Haldane model  \cite{owerreFirstTheoreticalRealization2016a,kimRealizationHaldaneKaneMeleModel2016}, the details of which we review in the Supplemental Material (SM) \footnote{See Supplemental Material for the details of the Haldane model and the effects of anisotropy, a full discussion on the particle-hole symmetry, details on the LLG simulations and transmission calculations, additional sources of disorder and an estimation of the energy scales. The Supplemental Material contains Refs.~\cite{fukuiChernNumbersDiscretized2005,castroAndersonLocalizationTopological2015}.}. In the magnon Haldane model, the Dzyaloshinskii-Moriya interaction (DMI) opens the topological gap.
	
	In the absence of dissipation, $\alpha=J_s=0$, we obtain two sets of two spin wave solutions to Eq.~(\ref{eq:bdg}), as a result of the particle-hole symmetry. However, this doubling is not a physical effect and merely the result of the fact that we represent the spin waves using complex scalar fields \cite{harmsAntimagnonics2022}. We can thus choose to only work with one branch of the solutions and we then obtain two bands with dispersion relations
	\begin{equation}
		\omega_{\bm k,1}^\pm = H_{\pm}+3JS+|\bm h_{\bm k}|,\quad \omega_{\bm k,2}^\pm=H_{\pm}+3JS-|\bm h_{\bm k}|,
	\end{equation} 
	where $J$ is the exchange constant and $\bm h_{\bm k}$ comprises the details of the magnon Haldane model \cite{Note2}. 
	We show this bulk dispersion in Fig.~\hyperref[fig:main-fig]{\ref*{fig:main-fig}(c-d)}, comparing equilibrium and non-equilibrium. In equilibrium, we obtain only states with positive frequencies, whereas in non-equilibrium, where $H=H_-<0$, the bands are shifted down in frequency, and we now obtain states with negative frequencies. The negative frequency modes have opposite handedness compared to the positive frequency modes, and thus rotate counterclockwise, whereas the positive frequency modes rotate clockwise. They also carry opposite angular momentum. The shift down in frequency can be explained from the fact that in non-equilibrium the effective magnetic field $\delta \mathcal{H}/\delta \bm S_{i}$ is pointing against the magnetization, thus lowering the frequency of the modes. 
	We refer the reader to a full discussion about the particle-hole symmetry and its implications to the SM \cite{Note2}, where we also discuss the stability and band structure in the presence of magnetic anisotropy.

	The topological invariant for this system, the Chern number of the band $n$, is now defined as $2\pi\Omega_{n}=\sum_{\bm k}\varepsilon_{ij}\partial_{k_i}\mathcal{A}_j^n$, where $	\mathcal{A}^n_j=i\bra*{\Psi_{\bm k}^n}\sigma_3\ket*{ \partial_{k_j}\Psi_{\bm k}^n}$ is the Berry connection \cite{shindouTopologicalChiralMagnonic2013, leinKreinSchrodingerFormalismBosonic2019,gunninkTheoryElectricalDetection2021} and $\Psi_{\bm k}^n$ is the $n$-th eigenstate. In the bulk band structure, Fig.~\hyperref[fig:main-fig]{\ref*{fig:main-fig}(c-d)}, we have indicated the Chern number, $\pm 1$, for the two bands. In equilibrium, the two bands have opposite Chern number and therefore there are topologically-protected chiral edge modes connecting the two bands. In non-equilibrium, the Chern number of the bands is preserved, and since one band is shifted down to negative frequency, we therefore expect the edge modes connecting the two bulk modes to cross zero frequency.
	
	To further illustrate the topological nature of the edge states, we show the bandstructure of a ribbon, 16 unit cells wide, with armchair edges in Fig.~\hyperref[fig:main-fig]{\ref*{fig:main-fig}(e-f)}, and indicate the edge localization in the colorscale. %In this work, 
	We have chosen compensated boundaries, such that the edge coordination number, i.e., the number of nearest neighbors, is equal to the bulk coordination number, and discuss the case of uncompensated boundaries in the SM \cite{Note2}. In equilibrium, we obtain topologically-protected edge states, as can be seen from their localization 
	and their dispersion crossing the bulk band gap, and they thus have a finite frequency. In non-equilibrium the edge states remain, but are lowered in frequency and in fact cross zero frequency. We still have one forward-moving mode localized on one side of the ribbon, and a backward-moving mode on the other side. However, there are forward- and backward-moving edge modes with both positive and negative frequencies, and thus opposite handedness.
	
\begin{figure}
		\includegraphics[width=\columnwidth]{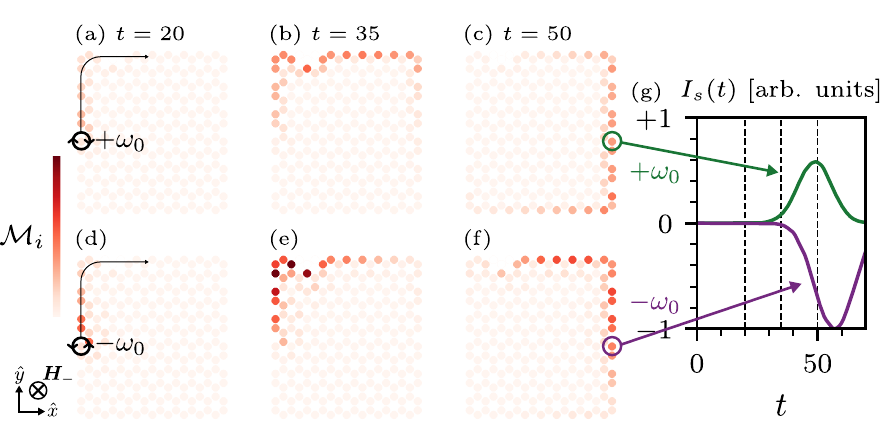}
		\caption{Spin dynamics simulation of a finite-size system in non-equilibrium, revealing the time evolution under a local excitation pulse with (a-c) positive and (d-f) negative frequency at $\pm\omega_0$, respectively. The chiral propagation direction of the edge modes is indicated by the arrow. The system starts in a uniform state, $\bm S = S\hat{\bm z}$, and is continuously excited at one single site at the left edge, circled in (a,d). (a-c) Snapshots of the time evolution of the spin-wave amplitude $\mathcal{M}_i(t)$ for a positive excitation frequency, $+\omega_0$. (d-f) Same as (a-c) but for a negative excitation frequency, $-\omega_0$. (g) The pumped spin current $I_s(t)$ for a site on the edge. The dashed vertical lines correspond to the times at which the snapshots in (a-f) are taken. 
			\label{fig:sim}}
	\end{figure}

	\textit{Numerical verification of the edge modes.} %In order 
	To verify the existence of the edge states at low frequencies, we numerically solve the LLG Eq.~\eqref{eq:LLG}, including Gilbert damping and the spin-orbit torque needed to stabilize the non-equilibrium setup. This allows us to capture the full dynamics, in particular nonlinearities that are not included in linear spin-wave theory. We describe the specifics of the simulations used in the SM \cite{Note2} and show the resulting dynamics in Fig.~\ref{fig:sim}. 
	We focus on the non-equilibrium dynamics, and use the same parameters used to calculate the bandstructure in Fig.~\hyperref[fig:main-fig]{\ref*{fig:main-fig}(c,f)}, and set $\alpha=10^{-3}$ and $J_s=\alpha H_-$. A spin wave is excited with positive frequency $\omega_0/JS=0.7$, Fig.~\hyperref[fig:sim]{\ref*{fig:sim}(a-c)}, and negative frequency $\omega_0/JS=-0.7$, Fig.~\hyperref[fig:sim]{\ref*{fig:sim}(d-f)}, at one single edge site using a transversely oscillating magnetic field with frequency $\omega_0$. We show the spin-wave amplitude, defined as the deviation of the spins from the $z$-axis, $\mathcal{M}_i(t) \equiv 1-S_i^z(t)$.
	
	For both positive and negative excitation frequency, an edge mode is excited, which travels clockwise around the system. Its topological stability due to the absence of backscattering is proven by its bypassing of the defect in the upper left corner, where three edge spins are missing. 
	Importantly, the excitations with opposite frequency have an opposite handedness, i.e., the individual spins rotate in the opposite direction in the $(x,y)$-plane and thus carry opposite angular momentum. This is in sharp contrast to the equilibrium situation, in which all magnons have the same chirality and 
	carry the same angular momentum. Experimentally, this difference could be accessed by means of time-resolved spin pumping \cite{chumakDirectDetectionMagnon2012}, because the pumped spin current $I_s(t) \propto \hat{\bm{z}} \cdot (\bm{S}_i(t) \times \dot{\bm{S}}_i(t))$ \cite{barkerThermalSpinDynamics2016} and the resulting spin-Hall voltage are a direct probe of handedness and, hence, of the topological negative-frequency excitations. To illustrate this, we show the pumped spin current, $I_s(t)$, for a site on the edge in Fig.~\hyperref[fig:sim]{\ref*{fig:sim}(g)}. %We observe that 
	$I_s(t)$ is opposite between positive and negative excitation frequencies, showing that the excitations carry opposite angular momentum. Moreover, the arrival times of opposite excitation frequencies differ, which we attribute to the different group velocity of the excited modes. This difference in group velocity can also be seen from the asymmetry of the band structure with respect to $\omega=0$, %as shown in
	Fig.~\hyperref[fig:main-fig]{\ref*{fig:main-fig}(f)}, and is tuneable by varying the %external 
	magnetic field. Finally, we observe that the absolute magnitude of $I_s(t)$ is larger for negative-frequency excitations. This is explained by the Gilbert damping, $\alpha\omega$, having the opposite sign for negative frequency modes compared to their positive frequency counterparts.

\begin{figure}
		
		\includegraphics[width=\columnwidth]{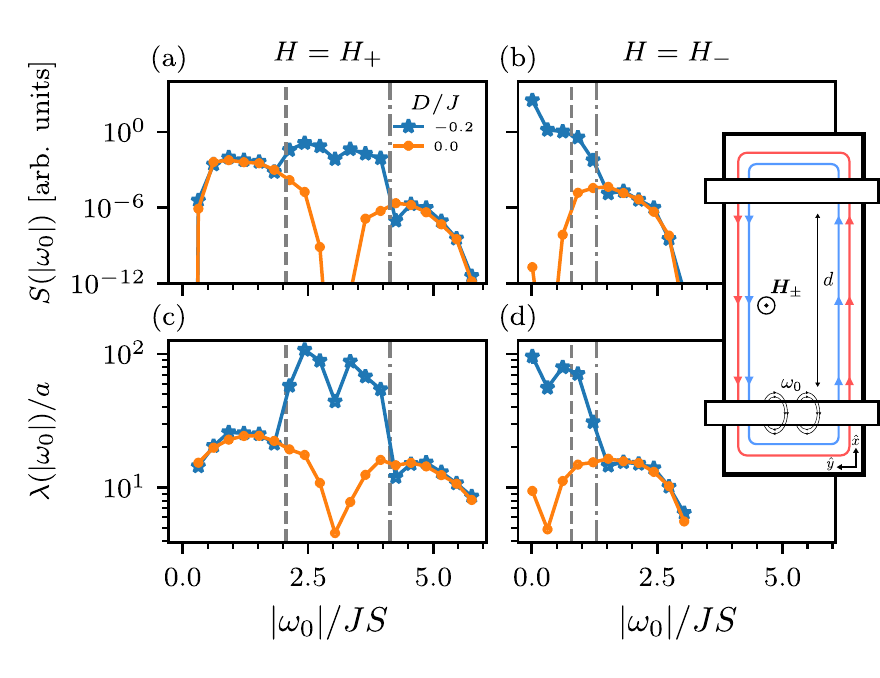}
		\caption{
			Propagating spin wave spectroscopy experiment as sketched in the inset, with edge modes excited by one antenna traveling through the film and picked up by the second antenna.
			(a-b) The transmission at finite disorder, $w=0.05$, and a fixed distance $d=200a$, as a function of excitation frequency $\omega$, for the equilibrium state ($H=H_+$) and the non-equilibrium state ($H=H_-$) for the topologically trivial state, $D=0$ and non-trivial state $D/J=-0.2$. Transmission is calculated with a finite Gilbert damping and a stabilizing spin-orbit torque for the non-equilibrium state. 
			(c-d) The corresponding decay length of the transmission. The dashed and dashed-dotted lines indicate the bottom and top of the bulk band gap.
			\label{fig:transport}}
	\end{figure}

	\textit{Propagating spin wave spectroscopy.} %One of the 
	A central goal in the field of magnon topology is the transport of angular momentum by topologically protected edge states in magnonic devices \cite{chumakMagnonSpintronics2015}. Since usual frequencies of the edge states are in the \si{THz} range, these cannot be excited using conventional microwave antennas. However, in the non-equilibrium setup,
	the edge states extend to zero frequency, and are therefore easily accessible. 
	We thus consider a propagating spin wave spectroscopy (PSWS) experiment \cite{vlaminckSpinwaveTransductionSubmicrometer2010}, where two antennas are placed a distance $d$ from each other. One antenna excites spin waves, 
	which are picked up by the second antenna after traveling through the film (see the inset of Fig.~\ref{fig:transport} for a device illustration). We consider a nanoribbon $10$ unit cells wide, with length $d$, orientated such that the edges are of the armchair type, which corresponds to the dispersion shown in Fig.~\hyperref[fig:main-fig]{\ref*{fig:main-fig}(e,f)}. %An antenna placed perpendicular to the propagation direction generates an Oersted field oscillating with frequency $\omega$ parallel to the propagation direction, and thus excites all possible spin waves with the frequencies $\pm \omega$.
	
	In the excitation antenna, the Oersted field oscillating with frequency $\omega$  excites all possible spin waves with the frequencies $\pm \omega$.
	Specifically, we model the excitation field by adding a local magnetic field term,  $\partial_t \bm S_i|_{\mathrm{exc}}=\bm S_i \times \bm h_i$, to the LLG Eq.~\eqref{eq:LLG}, expand in deviations %$m_{\pm}^i=(S_i^x\mp iS_i^y)/S$
	$m_i^\pm$, and numerically solve the resulting equation of motion to lowest nontrivial order in $m_i^\pm$ in position and frequency space. The second antenna is sensitive to the total microwave power, which we define as the transmission $S(\omega)\equiv\sum_{i\in \mathbb{R}_{p}}|m_i(\omega)|^2$, where $\mathbb R_p$ are the sites connected to the pickup antenna.  We also model a concentration $w$ of defects by removing spins, in order to capture the topological protection of the edge modes. 
	The details of this calculation are discussed in the SM \cite{Note2}, where we also consider three additional types of disorder to show that the robustness of the zero-frequency edge states is not dependent on the specific disorder considered in the main text.

	We show the resulting transmission in Fig.~\hyperref[fig:transport]{\ref*{fig:transport}(a-b)}, comparing the equilibrium %(a,c) 
	and non-equilibrium states, %(b,d) 
	and the topologically trivial state, $D=0$, and non-trival state $D/J=-0.2$, where $D$ is the strength of the DMI. We choose $\alpha=10^{-2}$ and stabilize the non-equilibrium state with a spin-orbit torque, $J_s=\alpha H$. 
	We first focus on the equilibrium state, $H_+/JS=0.1$, and simulate finite disorder, $w=0.05$, i.e., 5\% of all sites have a defect. 
	In Fig.~\hyperref[fig:transport]{\ref*{fig:transport}(a)}, we observe a broad peak in transmission at frequencies in the topologically nontrivial bulk band gap for $D/J=-0.2$. This feature is absent for $D=0$, proving that it is an effect of the non-trivial topology because the backscattering-immune edge states enable transmission while the bulk state transmission is suppressed. 
	
	Turning now to the topologically nontrivial non-equilibrium state, $H_-/JS=-2.75$ and $D/J=-0.2$, where the edge state lies around zero frequency [cf.~Fig.~\hyperref[fig:main-fig]{\ref*{fig:main-fig}(f)}], we see that transmission instead peaks around zero frequency. 
	Again, we find a clear distinction with the topologically trivial case, $D=0$, where transmission is suppressed at low frequencies at finite disorder. An important feature of the zero-frequency edge states is their higher transmission compared to the equilibrium edge states. This we attribute to the Gilbert damping, $\alpha\omega$, being proportional to frequency and thus lower for the zero-frequency edge states.
	
	The transmission follows an exponential decay as a function of distance, i.e., $S(d,\omega)\propto \exp(-d/\lambda(\omega))$, where the decay length $\lambda(\omega)$ is a function of the excitation frequency.
	We therefore fit the transmission over a range of separation distances $20a<d<200a$ ($a$ lattice constant) and obtain an estimate for the the decay length, $\lambda(\omega)$, which we show as a function of excitation frequency, $\omega$, in Fig.~\hyperref[fig:transport]{\ref*{fig:transport}(c-d)}. We observe that the decay length reflects the topological protection of the edge states, peaking when the edge states are excited. Furthermore, the decay lengths are much larger for $D/J=-0.2$ compared to $D=0$, reflecting the robustness against disorder of the edge states. Most importantly, in non-equilibrium, in the limit $\omega \rightarrow 0$ the decay length increases, which is in stark contrast with the equilibrium state, where the finite gap induced by the magnetic field blocks transmission.

	\textit{Discussion and Conclusion.} We have shown that by considering the magnetic excitations on top of a non-equilibrium state, stabilized by spin-orbit torques, we can effectively lower the frequency of topologically-protected chiral magnon edge modes. We obtain edge states with negative and positive frequency and we have confirmed their existence by numerically solving the Landau-Lifshitz-Gilbert equation, showing their stability and robustness against defects. Furthermore, we have shown that in a propagating spin wave spectroscopy experiment, the edge modes can be directly detected.
	
	In the SM we provide estimates for the required strength of the external magnetic field and spin-orbit torque for specific material choices \cite{Note2}. Here we note that in general the magnetic fields and spin-orbit torque, $\alpha H_-$, are proportional to the frequency of the edge mode in equilibrium. It would therefore be beneficial to consider this non-equilibrium state in a topological magnon crystal, where the frequencies of the edge modes is set by dipolar interaction, which is in the range of GHz \cite{shindouTopologicalChiralMagnonic2013,shindouChiralSpinwaveEdge2013}. An alternative approach would be to look at the transient regime, by first aligning to system to an external magnetic field, and then reversing the direction of the applied field. For a short transient period one would then observe the same features as discussed here, but after some time the system would relax to equilibrium.
	
	Our strategy can be used to lower other topological magnon excitations to zero frequency. Specifically, magnon Weyl semimetals would be an interesting prospect because zero-frequency Weyl points and associated topological surface states could come with the same transport anomalies as their finite-frequency counterparts \cite{suChiralAnomalyWeyl2017, mookTakingElectronmagnonDuality2018, liuMagnonQuantumAnomalies2019}. 
	Beyond magnons, it will be exciting to explore similar ideas for other bosonic Chern insulators, such as those formed by photons \cite{haldanePossibleRealizationDirectional2008} or phonons \cite{yangTopologicalAcoustics2015,wangTopologicalPhononicCrystals2015}. In these bosonic systems, non-equilibrium is accessible through external pumping, analogous to the spin-orbit torque used in this work.
	Finally, we note that non-equilibrium incoherent Hall-type transport \cite{onoseObservationMagnonHall2010,murakamiThermalHallEffect2017,kovalevSpinTorqueNernst2016} could be of interest because low-frequency edge states could potentially dominate transport.

	\begin{acknowledgments}
		R.A.D. is member of the D-ITP consortium, a program of the Dutch Organization for Scientific Research (NWO) that is funded by the Dutch Ministry of Education, Culture and Science (OCW). This work is
		in part funded by the Fluid Spintronics research programme with project number 182.069, financed by the Dutch Research
		Council (NWO), and by the Deutsche Forschungsgemeinschaft (DFG,
		German Research Foundation) - Project No.~504261060
		(Emmy Noether Programme).
	\end{acknowledgments}

	\onecolumngrid
\pagebreak

\begin{center}
	\textbf{\large Supplemental Material:\\
		Zero-frequency chiral magnonic edge states protected by non-equilibrium topology}
\end{center}

\makeatletter

\stepcounter{myequation}
\stepcounter{myfigure}
\stepcounter{mytable}
\renewcommand{\theequation}{S\arabic{equation}}
\renewcommand{\thefigure}{S\arabic{figure}}
\renewcommand{\bibnumfmt}[1]{[S#1]}

\section{Haldane model}
\begin{figure}[b]
	\centering
	\includegraphics[width=0.3\textwidth]{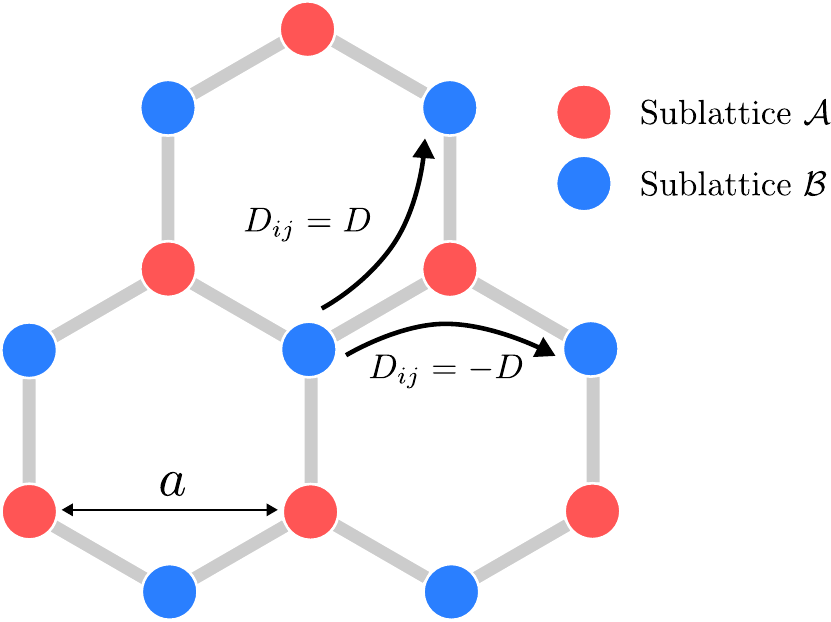}
	\caption{The honeycomb lattice of the Haldane model considered here. The relative sign of the Dzyaloshinskii-Moriya interaction is indicated. \label{fig:haldane}}
\end{figure}

In this work we consider the magnon Haldane model \cite{owerreFirstTheoreticalRealization2016a,kimRealizationHaldaneKaneMeleModel2016}, described by the two-dimensional Hamiltonian
\begin{equation}
	\mathcal H = -\frac{1}{2} \sum_{ij}[J_{ij} \bm S_i \cdot \bm S_j - D_{ij} \hat{\bm z} \cdot\left(\bm  S_i \times \bm S_j\right)] 
	-   \sum_i [H_{\pm}S_i^z - K_y (S_i^y)^2], 
	\label{eq:ham}
\end{equation}
where $\bm S_i$ are spins of length $S$ located on lattice sites $\bm R_i$ of a honeycomb lattice, as indicated in Fig.~\ref{fig:haldane}. Nearest neighbors experience an exchange coupling, $J_{ij}=J$, and next-nearest neighbors are coupled through the Dzyaloshinskii-Moriya interaction (DMI), $D_{ij}=-D_{ji}=D$. The spins are aligned to an external magnetic field, $H_{\pm}$, applied in the $z$ direction, contributing a Zeeman energy. We also consider an anisotropy with strength $K_y$, which will lead to elliptical precessions.

Furthermore, because we require a spin-orbit torque to stabilize the non-equilibrium state, there will be a current flowing in-plane in the heavy metal. This current will induce an Oersted field, affecting the spin dynamics. Assuming the current to flow along the $x$-direction, we have an additional torque in the LLG-equation~(\ref{eq:LLG}):
\begin{equation}
	\partial_t \bm S_i\big\rvert_{\mathrm{Oe}} = -\bm S_i \times H_{\mathrm{Oe}} \hat{\bm y},
\end{equation}
where $H_{\mathrm{Oe}}$ is the strength of the Oersted field. After linearization---which we will perform next---this torque can included in the effective spin-wave Hamiltonian.

We introduce the deviations $m^i_{\pm}=(S_i^x \mp i S_i^y)/\sqrt{2S}$ around the uniform state, $\bm S_i = S\hat{ \bm{z}}$, where $m^\pm_{i}$ refer to excitations for the equilibrium state ($H=H_+$) and non-equilibrium state ($H=H_-$). We expand the Hamiltonian, Eq.~(\ref{eq:ham}), up to the first nontrivial order in $m_\pm^i$, and obtain the quadratic spin-wave Hamiltonian
\begin{equation}
	\mathcal H_2 = \sum_{ij} \left[A_{ij} m_i^{\pm*} m_{j}^\pm + \frac{1}{2} B_{ij}\left( m_i^{\pm} m_{j}^{\pm} + m_i^{\pm*} m_{j}^{\pm*}\right)\right], \label{eq:ham2}
\end{equation}

%\begin{equation}
%	\bm{\mathcal{H}} = \begin{pmatrix}
	%		\bm m_{\pm}^* \bm m_{\pm} 
	%	\end{pmatrix}^T\begin{pmatrix}
	%		\bm{A} &  \bm B \\
	%		\bm B^\dagger & \bm A
	%	\end{pmatrix}
%	\begin{pmatrix}
	%		\bm m_\pm \\
	%		\bm m_{\pm}^* 
	%	\end{pmatrix}
%\end{equation}
where 
\begin{equation}
	A_{ij} =\delta_{ij}\left(H_{\pm}+H_{\mathrm{Oe}}+KS+S\sum_n J_{in}\right) - S(J_{ij} + iD_{ij});\quad 
	B_{ij} = \delta_{ij} (H_{\mathrm{Oe}}+K_yS), \label{eq:A}
\end{equation}
and we have included the current-induced Oersted contribution to the spin-wave dynamics.

We introduce the Fourier transform of the spin-wave operators, $m^\pm_{\mathcal{A/B},i} = \sqrt{2/N} \sum_{\bm k} e^{i\bm k\cdot \bm R_i}m_{\mathcal{A/B},\bm k}^\pm$ for the sublattices $\mathcal A/\mathcal B$ respectively, and obtain the Hamiltonian 
\begin{equation}
	\bm{\mathcal{H}}_{\bm k} =  \sum_{\bm k}  \bm \Psi_{\bm k}^{\pm^\dagger}\left[ (H_{\pm}+H_{\mathrm{Oe}}+K_yS+3JS)\sigma_0\tau_0 +(H_{\mathrm{Oe}}+K_yS)\sigma_0\tau_1 + (\bm h_{\bm k} \cdot \bm\sigma)\tau_0 \right] \bm \Psi_{\bm k}^\pm,
\end{equation}
where $\bm \Psi_{\bm k}^\pm = (m^\pm_{\mathcal{A},\bm k}, m^\pm_{\mathcal{B},\bm k,} m_{\mathcal{A},-\bm k}^{\pm*}, m^{\pm *}_{\mathcal{B},-\bm k} )^T$ is the magnon state vector, $\sigma_{\eta}$ are the Pauli matrices in the sublattice space, $\tau_{\eta}$ are the Pauli matrices in particle-hole space, $\bm \sigma$ is a pseudovector of Pauli matrices and
\begin{equation}
	\bm{h}_{\bm k} = S \sum_i 
	\begin{pmatrix}
		-J \cos(\bm{k} \cdot \bm{\delta}_i) \\
		J \sin(\bm{k} \cdot \bm{\delta}_i)\\
		2D \sin(\bm{k} \cdot \bm{\rho}_i)
	\end{pmatrix},
\end{equation}
where $\bm \delta_i$ and $\bm \rho_i$ are the vectors connecting nearest and next nearest neighbors,
\begin{align}
	\bm \delta &= [(0,-a/\sqrt{3}), (a/2, a/\sqrt{3}),(-a/2, a/\sqrt{3}) ]^T,\\
	\bm \rho &= [ (a,0), (-a/2,\sqrt{3}a/2),(-a/2,-\sqrt{3}a/2) ]^T.
\end{align}

The spin dynamics are now described by the LLG-equation~(\ref{eq:LLG}) in the main text, and after linearization we obtain the BdG-like equation, which has two sets of solutions. Choosing the solutions with positive norm, we find that (disregarding anisotropy and the Oersted field)
\begin{equation}
	\omega_{\bm k,1}^\pm = H_{\pm}+3JS+|\bm h_{\bm k}|,\quad \omega_{\bm k,2}^\pm=H_{\pm}+3JS-|\bm h_{\bm k}|.
\end{equation} 
In absence of DMI, the two bands touch at the Dirac points ($\bm K=(4\pi/3a)$, $\bm K'=(2\pi/3a, 2\pi/\sqrt{3}a)$), but a non-zero DMI opens up a gap at these points. The dispersion at the Dirac points is given by $\omega_{\bm K,1}^\pm=\omega_{\bm K',1}^\pm=H_{\pm}+3JS+ 3\sqrt{3}DS$ and $\omega_{\bm K,2}^\pm=\omega_{\bm K',2}^\pm=H_{\pm}+3JS- 3\sqrt{3}DS$. The opening of the gap implies a non-trivial topology, as can be shown by calculating the Chern number of the bands \cite{fukuiChernNumbersDiscretized2005}. This gap remains topological in non-equilibrium, as discussed in the main text.

\section{Anisotropy}
\label{sec:ani}
\begin{figure}
	\includegraphics{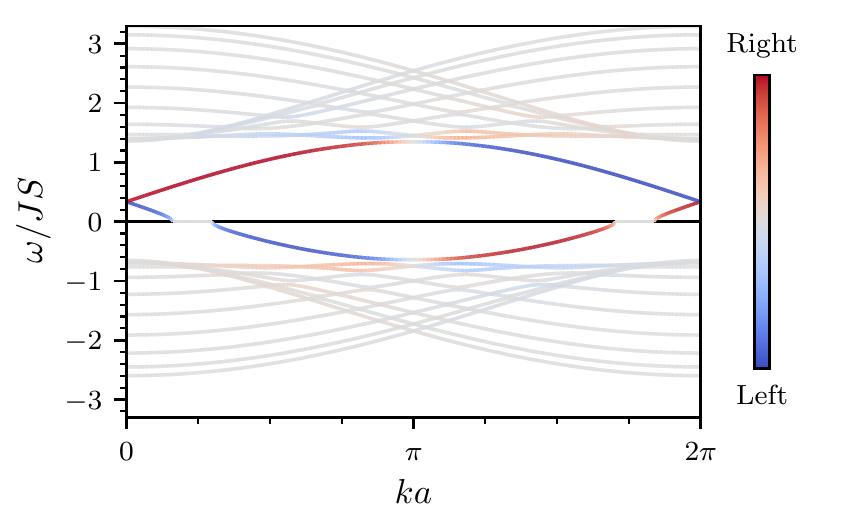}
	\caption{The magnon bandstructrure in the presence of anisotropy, $K/JS=0.1$. Where the edge states cross zero frequency we obtain an exceptional line. \label{fig:ani}}
\end{figure}
We now briefly discuss the effect of the effective anisotropy, $K\equiv H_{\mathrm{Oe}}+K_yS$, on the results obtained in the main text, where we have incorporated the effects of the current-induced Oersted field into the effective anisotropy. The anisotropy we consider yields elliptical precession, as can also be seen from the Hamiltonian, Eq.~(\ref{eq:ham}). Firstly, this has implications for the spin wave solutions up to first order in the dissipative terms, $\alpha$ and $J_s$, and up to zeroth order in $\bm k$, which now become
\begin{equation}
	\omega_{0,\pm} = \sqrt{(H_\pm+K)^2-K^2} - i [\alpha \left(H_\pm+K\right)-J_s].
\end{equation}
Therefore, the stability criterion is also changed and in order to have $\Im[\omega_{0,-}]<0$ in non-equilibrium, we require that 
\begin{equation}
	-J_s \ge \max[-\alpha(H_-+K), K].
\end{equation}
There are thus now two requirements: the spin-orbit torque has to overcome the Gilbert damping and the instability created by the anisotropy (giving elliptical precession). We note that in the Haldane model we consider here the magnetic field is orientated perpendicular to the plane, and there is therefore no ellipticity induced by the shape anisotropy. The only possible source of anisotropy is therefore the magnetocrystalline anisotropy, in addition to the anisotropy resulting from the current-induced Oersted field.

In the presence of anisotropy, we also observe the existence of an exceptional line where the edge states cross zero frequency, as shown in Fig.~\ref{fig:ani}. This exceptional line, a region in $\bm k$-space where $\Im[\omega_{\bm k}]\neq 0$  and $\Re[\omega_{\bm k}]=0$, is related to the same instability induced by anisotropy in the bulk system. It is thus present even in the absence of dissipation. However, it is a local instability at the edges of the system, and will therefore only lead to a local canting of the spins, which we have confirmed to be small with numerical LLG simulations. This canting will be counteracted by the applied spin-orbit torque, and therefore the system could always be stabilized with a strong enough spin-orbit torque. However, because the instability occurs at finite $\bm k$, it is not straight-forward to accurately determine the strength of the spin-orbit torque required. Finally, because of this canting the linearization procedure we apply is technically no longer valid, because we assume a uniform state to introduce the fluctuations. However, because we expect the canting to be small, we also expect the errors introduced by linearizing around a uniform state to be small. We therefore disregard the effects of anisotropy in the main text.

\section{Particle-hole symmetry}
\label{sec:phs}
As discussed in the main text, there are two sets of solutions to the BdG-like equation (\ref{eq:bdg}), due to the particle-hole symmetry (PHS), which implies that $\tau_x \bm ( H^\pm+i\alpha\omega^*\tau_z + i J_s\tau_0) \tau_x = - ( H^\pm -i\alpha\omega\tau_z + i J_s\tau_0)^*$. This implies that if $\omega$ is an eigenfrequency of $H^\pm$ with eigenvector $\bm \Psi=(u\ v)^T$ then $-\omega^*$ is an eigenfrequency with eigenvector $(v^*\ u^* )^T$. These two modes have opposite norm, defined as $\norm{\Psi}=\bra{\Psi}\sigma_z\ket{\Psi}$ \cite{harmsAntimagnonics2022}. Note that we have chosen the orientation of the deviations, $m_i^\pm$, dependent on the sign of the magnetic field in order to obtain this definition of the norm. 

Because of the doubling, it is sufficient to only consider one set of solutions. In the main text, we choose to only consider the set of solutions with positive norm, which means that out of equilibrium we naturally obtain negative frequencies. Alternatively, one can also choose to work with only positive frequencies, and both positive and negative norms. One can even include both positive and negative norms as well as frequencies, and take care of the double counting with a factor $\tfrac{1}{2}$. 

In equilibrium, one set of solutions has positive norm and only positive frequencies, and the second set with negative norm has only negative frequencies. Therefore, only considering the positive-norm modes in equilibrium is equivalent to only considering positive frequencies, as is common practice.
However, when we consider spin-wave excitations on top of the non-equilibrium state, there may be positive-norm states with negative frequency, and negative-norm states with positive frequency. Here we therefore have to consider the full frequency range, including negative frequencies.

Upon quantization of the excitations, one naturally obtains magnons and antimagnons, defined as having respectively a positive and negative product of frequency and norm. Similar to particles and holes, the antimagnons also carry opposite spin compared to the magnons. This thus implies that the negative frequency excitations as shown in the main text become antimagnons upon quantization, and carry negative angular momentum. Classically, which is what we consider in the main text, this corresponds to excitations having opposite chirality. For a further discussion on the magnon and antimagnons we refer the reader to Harms \textit{et al.} \cite{harmsAntimagnonics2022}.

\section{Uncompensated boundaries}
\begin{figure*}
	\centering
	%	\begin{subfigure}[b]{0.5\textwidth}
		%		\raisebox{2em}{\includegraphics[width=\textwidth]{magnon_bandstructure_no_exc}}
		%	\end{subfigure}%
	%	\begin{subfigure}[b]{0.5\textwidth}
		%		\includegraphics[width=\textwidth]{armchair_transmission_diffusion_length_w_0.05}
		%	\end{subfigure}
	\includegraphics[width=\textwidth]{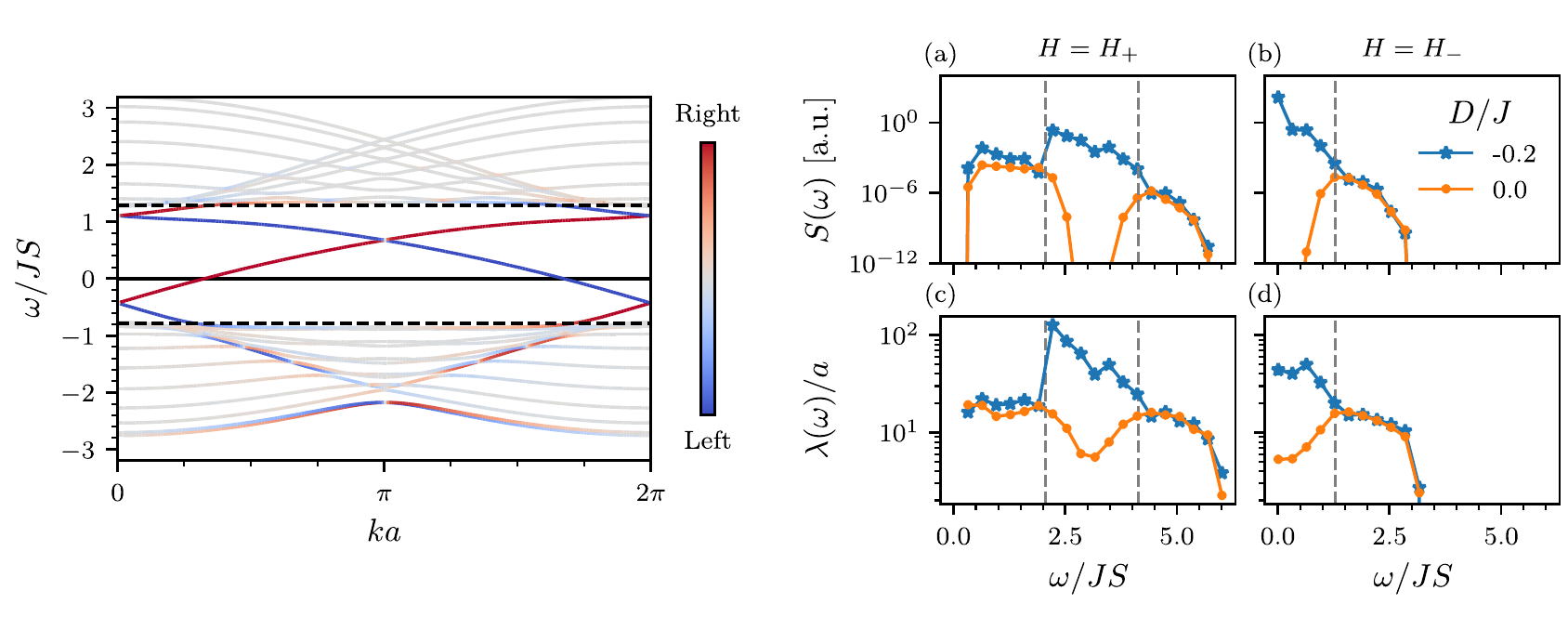}
	\caption{The nanoribbon bandstructure (left) and transmission and decay length (right) for uncompensated boundaries. \label{fig:uncompensated}}
\end{figure*}

\label{app:uncomp}
In this work, we have always considered compensated boundaries, such that the edge coordination number, i.e., the number of nearest neighbors, is equal to the bulk coordination number. This is implemented through substituting
\begin{equation}
	S\sum_n J_{in} \rightarrow 3SJ
\end{equation}
in the linearized Hamiltonian, Eq.~(\ref{eq:A}). In a real system this is not the case, and the effective field acting on edge sites will be lower. The edge excitations are thus lowered in frequency, but this does not affect the topological character of the edge modes. For completeness, we also show here the main results with uncompensated boundaries: the bandstructure for an armchair nanoribbon and the transmission and decay length, in Fig.~\ref{fig:uncompensated}. From the bandstructure we observe that the features as discussed in the main text are still present, with the edge modes crossing zero frequency. There are however small modifications to the specific dispersion of the edge mode, but from the transmission and decay length results we can conclude that these are insignificant. Furthermore, the main difference between the compensated and uncompensated boundaries appears for the equilibrium state, and the non-equilibrium state is barely affected, as can be seen from comparing Fig.~\ref{fig:transport} in the main text and Fig.~\ref{fig:uncompensated} in this supplementary material.

\section{Landau-Lifshitz-Gilbert simulations}
We consider a $10\times10$ unit cell structure and solve the Landau-Lifshitz-Gilbert (LLG) equation~(\ref{eq:LLG}) in the main text, and include 
%\begin{equation}
%	\partial_t \bm S_i = \bm S_i \times \left(-\frac{\partial \mathcal{H}}{\partial \bm S_i} + \bm h_i + \frac{\alpha}{S} \partial_t \bm S_i + \frac{J_s}{S}\bm S_i \times  \hat{\bm{z}}\right), \label{eq:LLG-transverse}
%\end{equation}
a transverse magnetic field, $\partial_t \bm S_i|_{\mathrm{exc}}=-\delta_{in}\, \bm S_i \times \bm h_i(t)$, where  
\begin{equation}
	h_i(t)=f_e(t)\,h_0\,(\cos(\omega_0 t), \sin(\omega_0 t),0)^T\,
\end{equation}
at one edge site $n$, with strength $h_0/S=10^{-6}$ and frequency $\omega_0/JS=0.7$. Here $f_e(t)=\exp(-(t-b)^2/2c^2)$ is an envelope function which slowly turns the pulse on and off, since turning on the excitation pulse instantaneously will excite a range of spurious frequencies. We choose $b=25$ and $c=10$.
We thus excite only a specific frequency $\omega_0$ and only one edge mode with a specific handedness. At $t=0$ the system is in the uniform state, $\bm S_i = S\hat z$. We set $D/J=-0.2$, such that we are in the topologically non-trivial regime. In non-equilibrium, there are edge modes close to zero frequency, and we thus expect to excite one of those edge modes. We also choose to work with compensated boundaries, and therefore apply a magnetic field of strength $JS$ to all edge sites with only two neighbors.

Since we also want to determine the stability of the system, we consider both Gilbert damping, $\alpha=10^{-3}$ and the spin-orbit torque $J_s=\alpha H_-$ needed to render this state stable. The snapshots in the main text, Fig.~\ref{fig:sim},  show the spin-wave amplitude, defined as the deviation of the spins from the $z$-axis, $\mathcal{M}_i(t) \equiv 1-S_i^z(t)$. The three missing atoms in the top-left corner are modeled by applying a large on-site magnetic field, rendering them effectively inaccessible to spin-waves.

%We take the first snapshot at $t=5$ and every $\Delta t=15$ after, resulting in the three snapshots as shown in Fig.~\ref{fig:sim} in the main text. The corresponding videos are \texttt{equilibrium.mp4}, for the equilibrium state (top row), \texttt{non-equilibrium.mp4} for the non-equilibrium state (middle row) and \texttt{non-equilibrium-defect.mp4} for the non-equilibrium state, including a defect (bottom row).

\section{Transmission}

We model the excitation field induced the nanoantenna by adding to the LLG equation~(\ref{eq:LLG}) in the main text, a local Oersted field oscillating with frequency $\omega$ parallel to the propagation
direction, $\partial_t \bm S_i|_{\mathrm{exc}} = \delta_{in} \bm S_i \times \bm h_{\mathrm{exc}}$, where $\bm h_{\mathrm{exc}}=b_0\cos \omega t\,\hat x$. We expand in deviations $m_{\pm}^i=(S_i^x\mp iS_i^y)/\sqrt{2S}$ and after Fourier transforming to frequency space we obtain the following equation of motion in position space,
\begin{equation}
	{\mathbb G} ^{-1} (\omega) \bm\Psi(\omega) = \bm h(\omega) \label{eq:eom}.
\end{equation}
Here $\bm\Psi(\omega)=[m_1,\dots,m_N,m_1^*,\dots,m_N^*]^T$ is the spin wave state vector, $\bm h(\omega)=[h_1(\omega),\dots,h_N(\omega),h_1^*(\omega),\dots,h_N^*(\omega)]$ is the Fourier transform of the circular components $h_i(\omega) = (h_i^x \mp ih_i^y)\delta_{i\in \mathbb{R}_{a}} $ of the excitation field, which is only non-zero for the sites $\mathbb{R}_{a}$ connected to the antenna. 
The inverse magnon propagator is given by \cite{ruckriegelBulkEdgeSpin2018}
\begin{equation}
	\mathbb G ^{-1} (\omega) = \tau_z  \sigma_0 \omega + \tau_0  \sigma_0(\alpha\omega-J_s)  - \tau_z  \bm{A},
\end{equation}
where $\sigma$ and $\tau$ are the Pauli matrices defined in the sublattice and magnon/antimagnon space respectively, and $\bm A$ was defined in Eq.~(\ref{eq:A}). Note here that the transmission is calculated after linearization in deviation from a uniform state, and therefore anisotropy has to be excluded, since this will lead to a finite canting of the spins, as discussed in Sec.~\ref{sec:ani}. However, we expect that after linearization around this canted state the inverse magnon propagator will be close to the one obtained after linearziation around a uniform state, and thus only a small error is introduced. 

Specifically, we are interested in the signal generated by the pickup antenna, which is sensitive to the total microwave power $S(\omega)=\sum_{i\in \mathbb{R}_{p}}|m_i(\omega)|^2$ of the sites $\mathbb R_p$ connected to the pick-up antenna. By using the solution of the linearized LLG (\ref{eq:eom}), $m_i(\omega)  = \sum_j \mathbb G_{ij}(\omega)h_j(\omega)$, we can write this as
\begin{equation}
	S(\omega) = \Tr_{i\in \mathbb{R}_{p}} \left[\mathbb{G}(\omega) \mathbb{H}(\omega)\mathbb G^\dagger(\omega) \right],
\end{equation}
where $\mathbb H(\omega)=\diag[|\bm h(\omega)|^2]$ and the trace is performed over the sites $\mathbb R_p$ which are connected to the pick-up antenna. Here it is important to note that the antenna is placed perpendicularly to the propagation direction, as shown in the inset of Fig.~\ref{fig:transport} in the main text. The Oersted field generated by the antenna will induce a transverse magnetic field, $\cos\omega t\,\hat x$, and thus will excite both positive and negative frequencies, as can be readily seen by taking the Fourier transform of this excitation field. Therefore, both positive and negative frequency spin waves are excited. In order to account for this, we have summed over both positive- and negative-norm solutions by tracing over the entire particle-hole space, for only positive frequency. This is equivalent to considering both positive and negative frequencies, and only the positive norm, as was also explained in Sec.~\ref{sec:phs}, and is done for numerical simplicity. Because both positive and negative frequencies are excited, we show the absolute frequency of the bulk band throughout this work. For excitation frequencies below the bottom of the absolute bulk gap two edge modes with positive and negative frequency can be excited, whilst between the bottom and top of the absolute bulk gap only one edge mode with positive frequency exists. 

We add a large on-site magnetic field to $wN$ randomly chosen lattice sites, where $N$ is the total number of lattice sites and $w\in[0,1]$ is the disorder concentration. This large magnetic field makes these sites effectively inaccessible for the spin waves. We average over multiple realizations of the disorder until we reach convergence. 

\begin{figure*}
	\centering
	\includegraphics{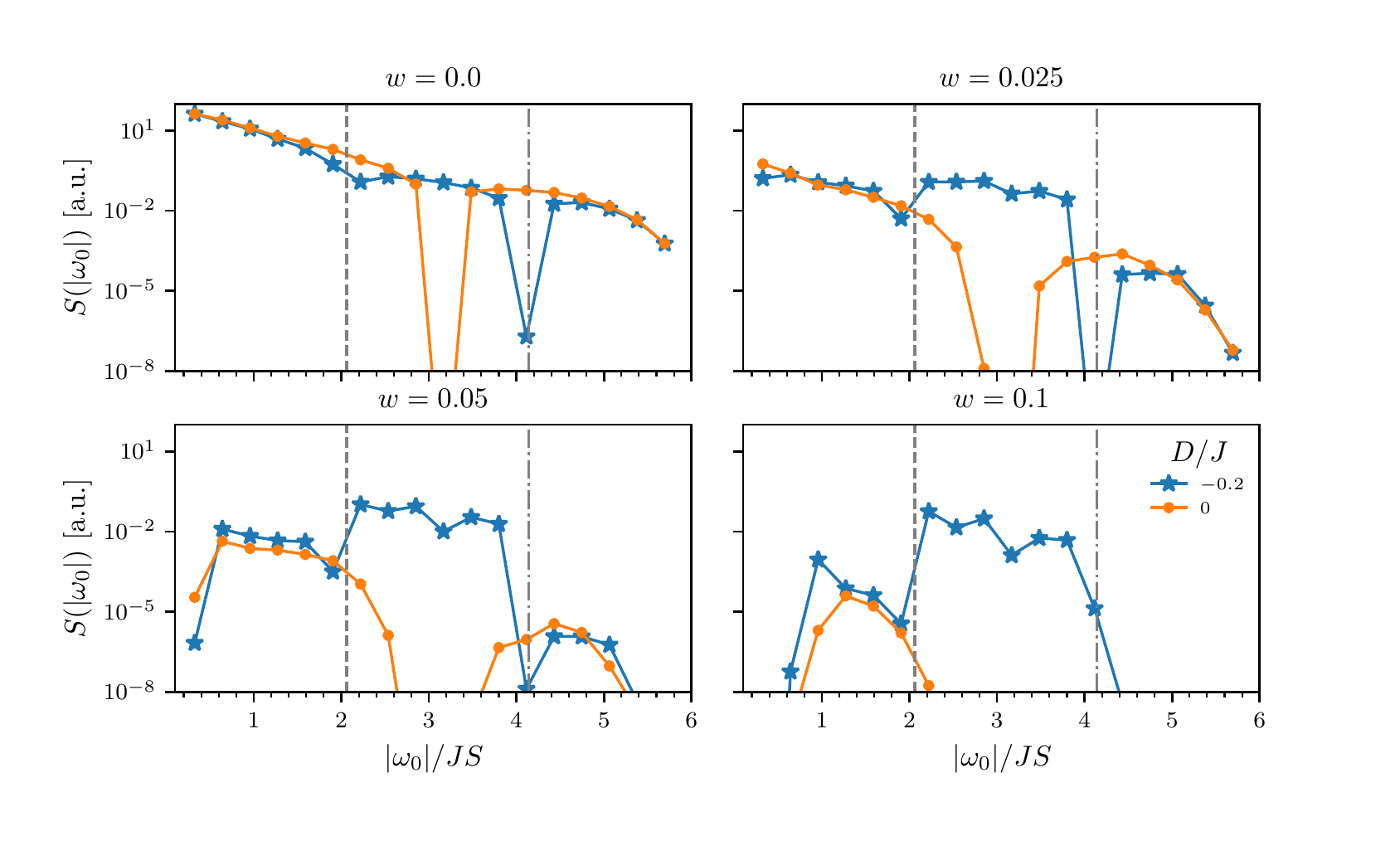}
	\caption{The transmission for increasing disorder $w$, as a function of excitation frequency, in equilibrium ($H_+/JS=0.1$), comparing the topologically trivial ($D=0$) and non-trivial ($D\neq0$) phases. The bottom and top of the bulk band gap is indicated by the dashed and dashed-dotted gray lines. The sharps dips in transmission, which are most prominent for zero disorder, are the result of the absence of modes with that specific frequency. \label{fig:poc}}
\end{figure*}
We will discuss here the transmission for the equilibrium state, in order to show that this formalism captures the topological protection of the edge modes. We show the transmission for increasing disorder in Fig.~\ref{fig:poc}. We can clearly see that for increasing disorder levels, the bulk modes are suppressed, while the edge modes, which lie in the bulk band gap, are unaffected. This can also be seen by comparing to the topologically trivial phase, where $D=0$, which does not contain such edge states, to the topologically non-trivial phase, $D/J=-0.2$. From the magnitude of the transmission of the edge states in the bulk band gap for increasing disorder, we also observe that the edge modes are barely affected by the increasing disorder level. This we can thus attribute to the topological protection of the edge modes, which disallows backscattering. We can therefore conclude that the transmission of spin waves in the bulk band gap, in presence of a finite disorder level, is a good indication of topologically protected edge modes.

We observe that there are fluctuations in the transmission signal as a function of frequency inside the gap in Fig.~\ref{fig:transport}. We propose that these fluctuations are related to a combination of the Gilbert damping and variations in the group velocity of the edge modes, combined with the presence of defects.
The  topological protection implies that the edge modes will travel around the defects, similar to the effect shown in Fig.~\ref{fig:sim}. Therefore, the total path traveled and thus the passage time of the modes increases also, which causes them to be damped out further by the Gilbert damping. There are further variations because the group velocity is not constant across the gap and therefore modes with different frequencies have different passage times. 
The dip in transmission in the middle of the band gap is most likely the result of the disorder creating a mid-gap impurity band in Chern insulators \cite{castroAndersonLocalizationTopological2015}.
We also note that the Gilbert damping scales as $\alpha\omega_0$, which reduces the signal for larger frequencies. We have furthermore confirmed that the results as shown in Fig.~\ref{fig:transport} have converged as a function of disorder ensemble size, and they are therefore not artifacts of the disorder sampling.

\section{Additional sources of disorder}
\begin{figure*}
	\centering
	\includegraphics{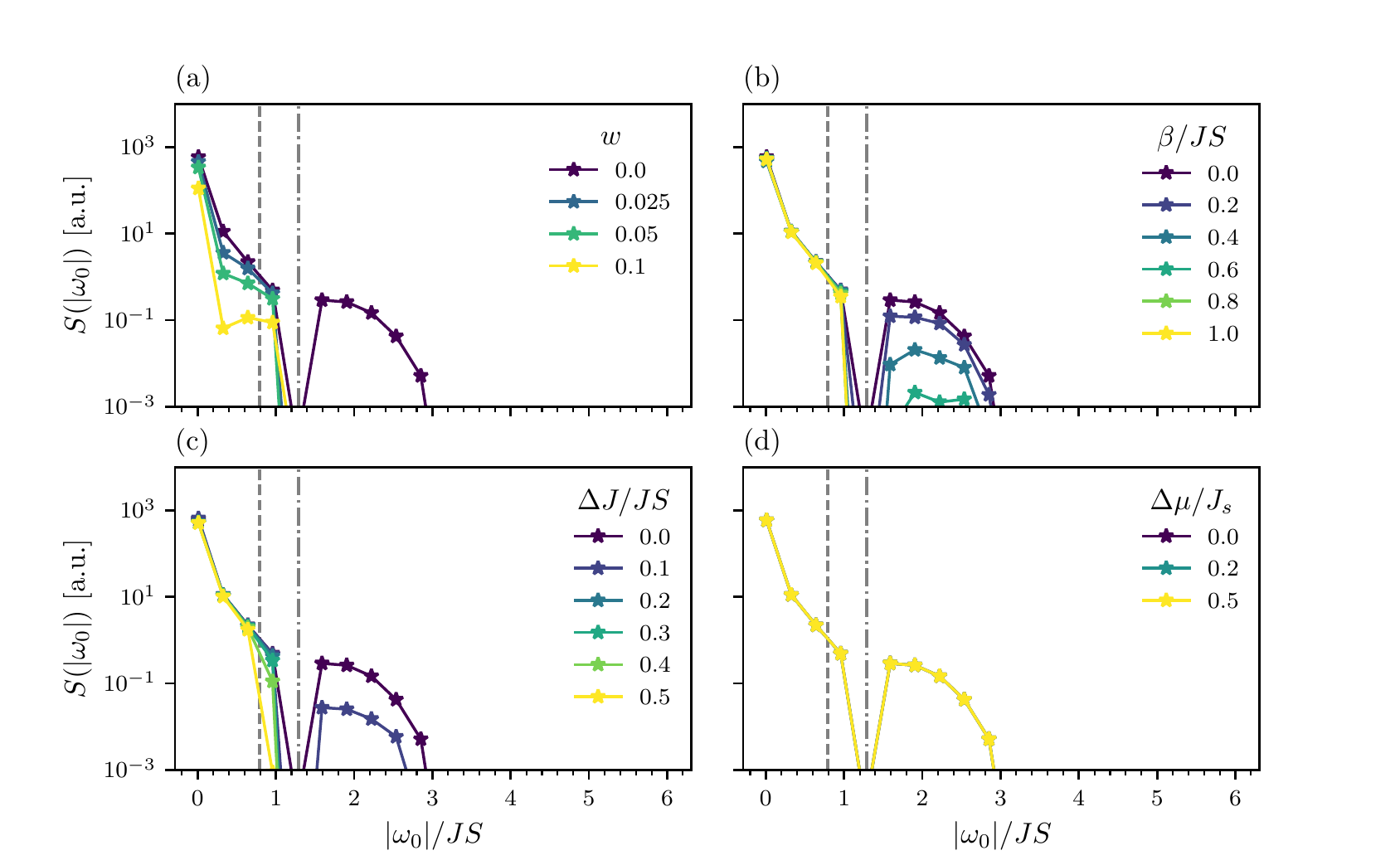}
	\caption{Transmission in the non-equilibrium state for four types of disorder: (a) defects with concentration $w$, (b) random on-site magnetic fields parameterized by the distribution width $\beta$, (c) fluctuating bonds parameterized by the distribution width $\Delta J$ and (d) non-uniform spin-orbit torque, parameterized by the distribution width $\Delta \mu$. The dashed and dashed-dotted lines indicate the bottom and top of the bulk band gap. \label{fig:disorder-study}}
\end{figure*}
In any real system there can be a wide variety of disorder. We therefore consider here three further sources of disorder, in addition to the defects as described in the main text: a random on-site magnetic potential, drawn from the uniform distribution $[-\frac\beta2,+\frac\beta2]$; fluctuating bond strengths, such that $J\rightarrow J+\delta J$, where $\delta J$ is drawn from the uniform distribution $[-\frac{\Delta J}{2},+\frac{\Delta J}{2}]$; and non-uniformity in the heavy metal\textbar ferromagnet interface, modeled by adding local fluctuations $\delta \mu$ to the applied spin-orbit torque drawn from the uniform distribution $[-\frac{\Delta \mu}{2},+\frac{\Delta \mu}{2}]$, such that $J_s \rightarrow J_s + \delta\mu$. 

We show the results for all four kinds of disorder in Fig.~\ref{fig:disorder-study} for the non-equilibrium state. A full study of the effects of disorder is beyond the scope of this Supplementary Material, but we can draw the important conclusion that the zero-frequency edge states in non-equilibrium can be measured through a propagating spin wave spectroscopy experiment, at moderate levels of disorder.

% the disorder induced by defects, Fig.~\hyperref[fig:disorder-study]{\ref*{fig:disorder-study}(a)}, behaves qualitatively different from the three additional sources of disorder, Fig.~\hyperref[fig:disorder-study]{\ref*{fig:disorder-study}(b-d)}. Most importantly, the transmission of the edge states is also affected by the defects, whilst it is unaffected by the random on-site magnetic field, random bond disorder and non-uniform spin-orbit torque.
%This we attribute to the fact that the edge states can also scatter elastically off the defects, reducing the lifetime of the edge states. 
%
%In the case of the three
%Furthermore, random on-site magnetic fields is know to not affect the transmission of edge states in the magnon Chern insulator in equilibrium \cite{wangBosonicBottIndex2020}.
%
%The non-uniform spin-orbit torque, Fig.~\hyperref[fig:disorder-study]{\ref*{fig:disorder-study}(c)}, does not affect the transmission, which we attribute to the fact that the spin-orbit torque does not play a role in the spin dynamics.\footnote{Note that in the linear spin-wave theory we do not include the effects of the current-induced Oersted field.} It can however dampen or amplify the spin-wave modes, but this is effect will vanish on average.

\section{Estimation of energy scales}
It is useful to consider here the energy scales of real materials, in order to determine what strength of magnetic field would be required to lower the frequency to a accessible level. In order to push the edge modes down to zero energy, we need to apply an opposite magnetic field equal to the bulk gap in the absence of a magnetic field. As a rule of thumb, a magnetic field of strength $H=\SI{1}{T}$ corresponds to an energy $g\mu_B H$ of approximately $\SI{0.1}{meV}\approx \SI{0.1}{THz}$ (assuming a g-factor of 2). 

The edge states in CrI\textsubscript{3}, a prominent topological magnon insulator candidate, have energies ${\sim}\SI{10}{meV}$ \cite{chenTopologicalSpinExcitations2018}, and would therefore require very large magnetic fields (\SI{100}{T}), making it not the best candidate to realize the zero-energy edge state as proposed here. Alternative candidates would be the kagome lattice ferromagnet Cu(1,3-bdc), where the topological excitations have energies ${\sim}\SI{1}{THz}$ \cite{chisnellTopologicalMagnonBands2015} or the YIG\textbar Fe magnonic crystal as proposed by Shindou \textit{et al.} \cite{shindouTopologicalChiralMagnonic2013}, where (depending on the specific implementation), the edge modes have energies ${\sim}\SI{35}{GHz}$. Even though these two systems are not realizations of the honeycomb Haldane model, the non-equilibrium state should still exhibit low energy topological edge modes, although the specific dispersion of the edge modes might be different.

The spin-orbit torque strength required to stabilize the non-equilibrium state is $J_s=\alpha H_-$. Therefore, when a strong magnetic field is required to lower the edge states to zero energy, the spin-orbit torque required is also large. The specific strength of the spin-orbit torque in real systems will depend on the specifics of the heavy metal\textbar ferromagnet bilayer system, such as the interface roughness, the spin Hall angle of the heavy metal used and the Gilbert damping of the ferromagnet. It is however clear that here it would also be beneficial to have as small a spin-orbit torque as possible, to prevent unwanted side-effects such as heating and structural deformations.


\begin{thebibliography}{57}%
	\makeatletter
	\providecommand \@ifxundefined [1]{%
		\@ifx{#1\undefined}
	}%
	\providecommand \@ifnum [1]{%
		\ifnum #1\expandafter \@firstoftwo
		\else \expandafter \@secondoftwo
		\fi
	}%
	\providecommand \@ifx [1]{%
		\ifx #1\expandafter \@firstoftwo
		\else \expandafter \@secondoftwo
		\fi
	}%
	\providecommand \natexlab [1]{#1}%
	\providecommand \enquote  [1]{``#1''}%
	\providecommand \bibnamefont  [1]{#1}%
	\providecommand \bibfnamefont [1]{#1}%
	\providecommand \citenamefont [1]{#1}%
	\providecommand \href@noop [0]{\@secondoftwo}%
	\providecommand \href [0]{\begingroup \@sanitize@url \@href}%
	\providecommand \@href[1]{\@@startlink{#1}\@@href}%
	\providecommand \@@href[1]{\endgroup#1\@@endlink}%
	\providecommand \@sanitize@url [0]{\catcode `\\12\catcode `\$12\catcode
		`\&12\catcode `\#12\catcode `\^12\catcode `\_12\catcode `\%12\relax}%
	\providecommand \@@startlink[1]{}%
	\providecommand \@@endlink[0]{}%
	\providecommand \url  [0]{\begingroup\@sanitize@url \@url }%
	\providecommand \@url [1]{\endgroup\@href {#1}{\urlprefix }}%
	\providecommand \urlprefix  [0]{URL }%
	\providecommand \Eprint [0]{\href }%
	\providecommand \doibase [0]{https://doi.org/}%
	\providecommand \selectlanguage [0]{\@gobble}%
	\providecommand \bibinfo  [0]{\@secondoftwo}%
	\providecommand \bibfield  [0]{\@secondoftwo}%
	\providecommand \translation [1]{[#1]}%
	\providecommand \BibitemOpen [0]{}%
	\providecommand \bibitemStop [0]{}%
	\providecommand \bibitemNoStop [0]{.\EOS\space}%
	\providecommand \EOS [0]{\spacefactor3000\relax}%
	\providecommand \BibitemShut  [1]{\csname bibitem#1\endcsname}%
	\let\auto@bib@innerbib\@empty
	%</preamble>
	\bibitem [{\citenamefont {Haldane}(1988)}]{haldaneModelQuantumHall1988}%
	\BibitemOpen
	\bibfield  {author} {\bibinfo {author} {\bibfnamefont {F.~D.~M.}\
			\bibnamefont {Haldane}},\ }\bibfield  {title} {\bibinfo {title} {Model for a
			{{Quantum Hall Effect}} without {{Landau Levels}}: {{Condensed-Matter
					Realization}} of the "{{Parity Anomaly}}"},\ }\href
	{https://doi.org/10.1103/PhysRevLett.61.2015} {\bibfield  {journal} {\bibinfo
			{journal} {Physical Review Letters}\ }\textbf {\bibinfo {volume} {61}},\
		\bibinfo {pages} {2015} (\bibinfo {year} {1988})}\BibitemShut {NoStop}%
	\bibitem [{\citenamefont {Hasan}\ and\ \citenamefont
		{Kane}(2010)}]{hasanColloquiumTopologicalInsulators2010}%
	\BibitemOpen
	\bibfield  {author} {\bibinfo {author} {\bibfnamefont {M.~Z.}\ \bibnamefont
			{Hasan}}\ and\ \bibinfo {author} {\bibfnamefont {C.~L.}\ \bibnamefont
			{Kane}},\ }\bibfield  {title} {\bibinfo {title} {Colloquium: {{Topological}}
			insulators},\ }\href {https://doi.org/10.1103/RevModPhys.82.3045} {\bibfield
		{journal} {\bibinfo  {journal} {Reviews of Modern Physics}\ }\textbf
		{\bibinfo {volume} {82}},\ \bibinfo {pages} {3045} (\bibinfo {year}
		{2010})}\BibitemShut {NoStop}%
	\bibitem [{\citenamefont {Haldane}\ and\ \citenamefont
		{Raghu}(2008)}]{haldanePossibleRealizationDirectional2008}%
	\BibitemOpen
	\bibfield  {author} {\bibinfo {author} {\bibfnamefont {F.~D.~M.}\
			\bibnamefont {Haldane}}\ and\ \bibinfo {author} {\bibfnamefont
			{S.}~\bibnamefont {Raghu}},\ }\bibfield  {title} {\bibinfo {title} {Possible
			{{Realization}} of {{Directional Optical Waveguides}} in {{Photonic
					Crystals}} with {{Broken Time-Reversal Symmetry}}},\ }\href
	{https://doi.org/10.1103/PhysRevLett.100.013904} {\bibfield  {journal}
		{\bibinfo  {journal} {Physical Review Letters}\ }\textbf {\bibinfo {volume}
			{100}},\ \bibinfo {pages} {013904} (\bibinfo {year} {2008})}\BibitemShut
	{NoStop}%
	\bibitem [{\citenamefont {Ozawa}\ \emph {et~al.}(2019)\citenamefont {Ozawa},
		\citenamefont {Price}, \citenamefont {Amo}, \citenamefont {Goldman},
		\citenamefont {Hafezi}, \citenamefont {Lu}, \citenamefont {Rechtsman},
		\citenamefont {Schuster}, \citenamefont {Simon}, \citenamefont {Zilberberg},\
		and\ \citenamefont {Carusotto}}]{ozawaTopologicalPhotonics2019a}%
	\BibitemOpen
	\bibfield  {author} {\bibinfo {author} {\bibfnamefont {T.}~\bibnamefont
			{Ozawa}}, \bibinfo {author} {\bibfnamefont {H.~M.}\ \bibnamefont {Price}},
		\bibinfo {author} {\bibfnamefont {A.}~\bibnamefont {Amo}}, \bibinfo {author}
		{\bibfnamefont {N.}~\bibnamefont {Goldman}}, \bibinfo {author} {\bibfnamefont
			{M.}~\bibnamefont {Hafezi}}, \bibinfo {author} {\bibfnamefont
			{L.}~\bibnamefont {Lu}}, \bibinfo {author} {\bibfnamefont {M.~C.}\
			\bibnamefont {Rechtsman}}, \bibinfo {author} {\bibfnamefont {D.}~\bibnamefont
			{Schuster}}, \bibinfo {author} {\bibfnamefont {J.}~\bibnamefont {Simon}},
		\bibinfo {author} {\bibfnamefont {O.}~\bibnamefont {Zilberberg}},\ and\
		\bibinfo {author} {\bibfnamefont {I.}~\bibnamefont {Carusotto}},\ }\bibfield
	{title} {\bibinfo {title} {Topological photonics},\ }\href
	{https://doi.org/10.1103/RevModPhys.91.015006} {\bibfield  {journal}
		{\bibinfo  {journal} {Reviews of Modern Physics}\ }\textbf {\bibinfo {volume}
			{91}},\ \bibinfo {pages} {015006} (\bibinfo {year} {2019})}\BibitemShut
	{NoStop}%
	\bibitem [{\citenamefont {Ma}\ \emph {et~al.}(2019)\citenamefont {Ma},
		\citenamefont {Xiao},\ and\ \citenamefont
		{Chan}}]{maTopologicalPhasesAcoustic2019}%
	\BibitemOpen
	\bibfield  {author} {\bibinfo {author} {\bibfnamefont {G.}~\bibnamefont
			{Ma}}, \bibinfo {author} {\bibfnamefont {M.}~\bibnamefont {Xiao}},\ and\
		\bibinfo {author} {\bibfnamefont {C.~T.}\ \bibnamefont {Chan}},\ }\bibfield
	{title} {\bibinfo {title} {Topological phases in acoustic and mechanical
			systems},\ }\href {https://doi.org/10.1038/s42254-019-0030-x} {\bibfield
		{journal} {\bibinfo  {journal} {Nature Reviews Physics}\ }\textbf {\bibinfo
			{volume} {1}},\ \bibinfo {pages} {281} (\bibinfo {year} {2019})}\BibitemShut
	{NoStop}%
	\bibitem [{\citenamefont {Jin}\ \emph {et~al.}(2016)\citenamefont {Jin},
		\citenamefont {Lu}, \citenamefont {Wang}, \citenamefont {Fang}, \citenamefont
		{Joannopoulos}, \citenamefont {Solja{\v c}i{\'c}}, \citenamefont {Fu},\ and\
		\citenamefont {Fang}}]{jinTopologicalMagnetoplasmon2016}%
	\BibitemOpen
	\bibfield  {author} {\bibinfo {author} {\bibfnamefont {D.}~\bibnamefont
			{Jin}}, \bibinfo {author} {\bibfnamefont {L.}~\bibnamefont {Lu}}, \bibinfo
		{author} {\bibfnamefont {Z.}~\bibnamefont {Wang}}, \bibinfo {author}
		{\bibfnamefont {C.}~\bibnamefont {Fang}}, \bibinfo {author} {\bibfnamefont
			{J.~D.}\ \bibnamefont {Joannopoulos}}, \bibinfo {author} {\bibfnamefont
			{M.}~\bibnamefont {Solja{\v c}i{\'c}}}, \bibinfo {author} {\bibfnamefont
			{L.}~\bibnamefont {Fu}},\ and\ \bibinfo {author} {\bibfnamefont {N.~X.}\
			\bibnamefont {Fang}},\ }\bibfield  {title} {\bibinfo {title} {Topological
			magnetoplasmon},\ }\href {https://doi.org/10.1038/ncomms13486} {\bibfield
		{journal} {\bibinfo  {journal} {Nature Communications}\ }\textbf {\bibinfo
			{volume} {7}},\ \bibinfo {pages} {13486} (\bibinfo {year}
		{2016})}\BibitemShut {NoStop}%
	\bibitem [{\citenamefont {Jin}\ \emph {et~al.}(2017)\citenamefont {Jin},
		\citenamefont {Christensen}, \citenamefont {Solja{\v c}i{\'c}}, \citenamefont
		{Fang}, \citenamefont {Lu},\ and\ \citenamefont
		{Zhang}}]{jinInfraredTopologicalPlasmons2017}%
	\BibitemOpen
	\bibfield  {author} {\bibinfo {author} {\bibfnamefont {D.}~\bibnamefont
			{Jin}}, \bibinfo {author} {\bibfnamefont {T.}~\bibnamefont {Christensen}},
		\bibinfo {author} {\bibfnamefont {M.}~\bibnamefont {Solja{\v c}i{\'c}}},
		\bibinfo {author} {\bibfnamefont {N.~X.}\ \bibnamefont {Fang}}, \bibinfo
		{author} {\bibfnamefont {L.}~\bibnamefont {Lu}},\ and\ \bibinfo {author}
		{\bibfnamefont {X.}~\bibnamefont {Zhang}},\ }\bibfield  {title} {\bibinfo
		{title} {Infrared {{Topological Plasmons}} in {{Graphene}}},\ }\href
	{https://doi.org/10.1103/PhysRevLett.118.245301} {\bibfield  {journal}
		{\bibinfo  {journal} {Physical Review Letters}\ }\textbf {\bibinfo {volume}
			{118}},\ \bibinfo {pages} {245301} (\bibinfo {year} {2017})}\BibitemShut
	{NoStop}%
	\bibitem [{\citenamefont
		{McClarty}(2022)}]{mcclartyTopologicalMagnonsReview2022}%
	\BibitemOpen
	\bibfield  {author} {\bibinfo {author} {\bibfnamefont {P.~A.}\ \bibnamefont
			{McClarty}},\ }\bibfield  {title} {\bibinfo {title} {Topological {{Magnons}}:
			{{A Review}}},\ }\href
	{https://doi.org/10.1146/annurev-conmatphys-031620-104715} {\bibfield
		{journal} {\bibinfo  {journal} {Annual Review of Condensed Matter Physics}\
		}\textbf {\bibinfo {volume} {13}},\ \bibinfo {pages} {171} (\bibinfo {year}
		{2022})}\BibitemShut {NoStop}%
	\bibitem [{\citenamefont {Katsura}\ \emph {et~al.}(2010)\citenamefont
		{Katsura}, \citenamefont {Nagaosa},\ and\ \citenamefont
		{Lee}}]{katsuraTheoryThermalHall2010}%
	\BibitemOpen
	\bibfield  {author} {\bibinfo {author} {\bibfnamefont {H.}~\bibnamefont
			{Katsura}}, \bibinfo {author} {\bibfnamefont {N.}~\bibnamefont {Nagaosa}},\
		and\ \bibinfo {author} {\bibfnamefont {P.~A.}\ \bibnamefont {Lee}},\
	}\bibfield  {title} {\bibinfo {title} {Theory of the {{Thermal Hall Effect}}
			in {{Quantum Magnets}}},\ }\href
	{https://doi.org/10.1103/PhysRevLett.104.066403} {\bibfield  {journal}
		{\bibinfo  {journal} {Physical Review Letters}\ }\textbf {\bibinfo {volume}
			{104}},\ \bibinfo {pages} {066403} (\bibinfo {year} {2010})}\BibitemShut
	{NoStop}%
	\bibitem [{\citenamefont {{van Hoogdalem}}\ \emph {et~al.}(2013)\citenamefont
		{{van Hoogdalem}}, \citenamefont {Tserkovnyak},\ and\ \citenamefont
		{Loss}}]{vanhoogdalemMagneticTextureinducedThermal2013}%
	\BibitemOpen
	\bibfield  {author} {\bibinfo {author} {\bibfnamefont {K.~A.}\ \bibnamefont
			{{van Hoogdalem}}}, \bibinfo {author} {\bibfnamefont {Y.}~\bibnamefont
			{Tserkovnyak}},\ and\ \bibinfo {author} {\bibfnamefont {D.}~\bibnamefont
			{Loss}},\ }\bibfield  {title} {\bibinfo {title} {Magnetic texture-induced
			thermal {{Hall}} effects},\ }\href
	{https://doi.org/10.1103/PhysRevB.87.024402} {\bibfield  {journal} {\bibinfo
			{journal} {Physical Review B}\ }\textbf {\bibinfo {volume} {87}},\ \bibinfo
		{pages} {024402} (\bibinfo {year} {2013})}\BibitemShut {NoStop}%
	\bibitem [{\citenamefont {Shindou}\ \emph
		{et~al.}(2013{\natexlab{a}})\citenamefont {Shindou}, \citenamefont
		{Matsumoto}, \citenamefont {Murakami},\ and\ \citenamefont
		{Ohe}}]{shindouTopologicalChiralMagnonic2013}%
	\BibitemOpen
	\bibfield  {author} {\bibinfo {author} {\bibfnamefont {R.}~\bibnamefont
			{Shindou}}, \bibinfo {author} {\bibfnamefont {R.}~\bibnamefont {Matsumoto}},
		\bibinfo {author} {\bibfnamefont {S.}~\bibnamefont {Murakami}},\ and\
		\bibinfo {author} {\bibfnamefont {J.-i.}\ \bibnamefont {Ohe}},\ }\bibfield
	{title} {\bibinfo {title} {Topological chiral magnonic edge mode in a
			magnonic crystal},\ }\href {https://doi.org/10.1103/PhysRevB.87.174427}
	{\bibfield  {journal} {\bibinfo  {journal} {Physical Review B}\ }\textbf
		{\bibinfo {volume} {87}},\ \bibinfo {pages} {174427} (\bibinfo {year}
		{2013}{\natexlab{a}})}\BibitemShut {NoStop}%
	\bibitem [{\citenamefont {Zhang}\ \emph {et~al.}(2013)\citenamefont {Zhang},
		\citenamefont {Ren}, \citenamefont {Wang},\ and\ \citenamefont
		{Li}}]{zhangTopologicalMagnonInsulator2013}%
	\BibitemOpen
	\bibfield  {author} {\bibinfo {author} {\bibfnamefont {L.}~\bibnamefont
			{Zhang}}, \bibinfo {author} {\bibfnamefont {J.}~\bibnamefont {Ren}}, \bibinfo
		{author} {\bibfnamefont {J.-S.}\ \bibnamefont {Wang}},\ and\ \bibinfo
		{author} {\bibfnamefont {B.}~\bibnamefont {Li}},\ }\bibfield  {title}
	{\bibinfo {title} {Topological magnon insulator in insulating ferromagnet},\
	}\href {https://doi.org/10.1103/PhysRevB.87.144101} {\bibfield  {journal}
		{\bibinfo  {journal} {Physical Review B}\ }\textbf {\bibinfo {volume} {87}},\
		\bibinfo {pages} {144101} (\bibinfo {year} {2013})}\BibitemShut {NoStop}%
	\bibitem [{\citenamefont {Mook}\ \emph {et~al.}(2014)\citenamefont {Mook},
		\citenamefont {Henk},\ and\ \citenamefont
		{Mertig}}]{mookEdgeStatesTopological2014}%
	\BibitemOpen
	\bibfield  {author} {\bibinfo {author} {\bibfnamefont {A.}~\bibnamefont
			{Mook}}, \bibinfo {author} {\bibfnamefont {J.}~\bibnamefont {Henk}},\ and\
		\bibinfo {author} {\bibfnamefont {I.}~\bibnamefont {Mertig}},\ }\bibfield
	{title} {\bibinfo {title} {Edge states in topological magnon insulators},\
	}\href {https://doi.org/10.1103/PhysRevB.90.024412} {\bibfield  {journal}
		{\bibinfo  {journal} {Physical Review B}\ }\textbf {\bibinfo {volume} {90}},\
		\bibinfo {pages} {024412} (\bibinfo {year} {2014})}\BibitemShut {NoStop}%
	\bibitem [{\citenamefont
		{Owerre}(2016)}]{owerreFirstTheoreticalRealization2016a}%
	\BibitemOpen
	\bibfield  {author} {\bibinfo {author} {\bibfnamefont {S.~A.}\ \bibnamefont
			{Owerre}},\ }\bibfield  {title} {\bibinfo {title} {A first theoretical
			realization of honeycomb topological magnon insulator},\ }\href
	{https://doi.org/10.1088/0953-8984/28/38/386001} {\bibfield  {journal}
		{\bibinfo  {journal} {Journal of Physics: Condensed Matter}\ }\textbf
		{\bibinfo {volume} {28}},\ \bibinfo {pages} {386001} (\bibinfo {year}
		{2016})}\BibitemShut {NoStop}%
	\bibitem [{\citenamefont {Kim}\ \emph {et~al.}(2016)\citenamefont {Kim},
		\citenamefont {Ochoa}, \citenamefont {Zarzuela},\ and\ \citenamefont
		{Tserkovnyak}}]{kimRealizationHaldaneKaneMeleModel2016}%
	\BibitemOpen
	\bibfield  {author} {\bibinfo {author} {\bibfnamefont {S.~K.}\ \bibnamefont
			{Kim}}, \bibinfo {author} {\bibfnamefont {H.}~\bibnamefont {Ochoa}}, \bibinfo
		{author} {\bibfnamefont {R.}~\bibnamefont {Zarzuela}},\ and\ \bibinfo
		{author} {\bibfnamefont {Y.}~\bibnamefont {Tserkovnyak}},\ }\bibfield
	{title} {\bibinfo {title} {Realization of the {{Haldane-Kane-Mele Model}} in
			a {{System}} of {{Localized Spins}}},\ }\href
	{https://doi.org/10.1103/PhysRevLett.117.227201} {\bibfield  {journal}
		{\bibinfo  {journal} {Physical Review Letters}\ }\textbf {\bibinfo {volume}
			{117}},\ \bibinfo {pages} {227201} (\bibinfo {year} {2016})}\BibitemShut
	{NoStop}%
	\bibitem [{\citenamefont {Mook}\ \emph
		{et~al.}(2021{\natexlab{a}})\citenamefont {Mook}, \citenamefont {Plekhanov},
		\citenamefont {Klinovaja},\ and\ \citenamefont
		{Loss}}]{mookInteractionStabilizedTopologicalMagnon2021}%
	\BibitemOpen
	\bibfield  {author} {\bibinfo {author} {\bibfnamefont {A.}~\bibnamefont
			{Mook}}, \bibinfo {author} {\bibfnamefont {K.}~\bibnamefont {Plekhanov}},
		\bibinfo {author} {\bibfnamefont {J.}~\bibnamefont {Klinovaja}},\ and\
		\bibinfo {author} {\bibfnamefont {D.}~\bibnamefont {Loss}},\ }\bibfield
	{title} {\bibinfo {title} {Interaction-{{Stabilized Topological Magnon
					Insulator}} in {{Ferromagnets}}},\ }\href
	{https://doi.org/10.1103/PhysRevX.11.021061} {\bibfield  {journal} {\bibinfo
			{journal} {Physical Review X}\ }\textbf {\bibinfo {volume} {11}},\ \bibinfo
		{pages} {021061} (\bibinfo {year} {2021}{\natexlab{a}})}\BibitemShut
	{NoStop}%
	\bibitem [{\citenamefont {Nakata}\ \emph {et~al.}(2017)\citenamefont {Nakata},
		\citenamefont {Kim}, \citenamefont {Klinovaja},\ and\ \citenamefont
		{Loss}}]{nakataMagnonicTopologicalInsulators2017}%
	\BibitemOpen
	\bibfield  {author} {\bibinfo {author} {\bibfnamefont {K.}~\bibnamefont
			{Nakata}}, \bibinfo {author} {\bibfnamefont {S.~K.}\ \bibnamefont {Kim}},
		\bibinfo {author} {\bibfnamefont {J.}~\bibnamefont {Klinovaja}},\ and\
		\bibinfo {author} {\bibfnamefont {D.}~\bibnamefont {Loss}},\ }\bibfield
	{title} {\bibinfo {title} {Magnonic topological insulators in
			antiferromagnets},\ }\href {https://doi.org/10.1103/PhysRevB.96.224414}
	{\bibfield  {journal} {\bibinfo  {journal} {Physical Review B}\ }\textbf
		{\bibinfo {volume} {96}},\ \bibinfo {pages} {224414} (\bibinfo {year}
		{2017})}\BibitemShut {NoStop}%
	\bibitem [{\citenamefont {Mook}\ \emph {et~al.}(2018)\citenamefont {Mook},
		\citenamefont {G{\"o}bel}, \citenamefont {Henk},\ and\ \citenamefont
		{Mertig}}]{mookTakingElectronmagnonDuality2018}%
	\BibitemOpen
	\bibfield  {author} {\bibinfo {author} {\bibfnamefont {A.}~\bibnamefont
			{Mook}}, \bibinfo {author} {\bibfnamefont {B.}~\bibnamefont {G{\"o}bel}},
		\bibinfo {author} {\bibfnamefont {J.}~\bibnamefont {Henk}},\ and\ \bibinfo
		{author} {\bibfnamefont {I.}~\bibnamefont {Mertig}},\ }\bibfield  {title}
	{\bibinfo {title} {Taking an electron-magnon duality shortcut from electron
			to magnon transport},\ }\href {https://doi.org/10.1103/PhysRevB.97.140401}
	{\bibfield  {journal} {\bibinfo  {journal} {Physical Review B}\ }\textbf
		{\bibinfo {volume} {97}},\ \bibinfo {pages} {140401(R)} (\bibinfo {year}
		{2018})}\BibitemShut {NoStop}%
	\bibitem [{\citenamefont {Kondo}\ \emph {et~al.}(2019)\citenamefont {Kondo},
		\citenamefont {Akagi},\ and\ \citenamefont
		{Katsura}}]{kondoMathbbZTopological2019}%
	\BibitemOpen
	\bibfield  {author} {\bibinfo {author} {\bibfnamefont {H.}~\bibnamefont
			{Kondo}}, \bibinfo {author} {\bibfnamefont {Y.}~\bibnamefont {Akagi}},\ and\
		\bibinfo {author} {\bibfnamefont {H.}~\bibnamefont {Katsura}},\ }\bibfield
	{title} {\bibinfo {title} {${\mathbb z}_2$ topological {{Invariant}} for
			{{Magnon Spin Hall Systems}}},\ }\href
	{https://doi.org/10.1103/PhysRevB.99.041110} {\bibfield  {journal} {\bibinfo
			{journal} {Physical Review B}\ }\textbf {\bibinfo {volume} {99}},\ \bibinfo
		{pages} {041110(R)} (\bibinfo {year} {2019})}\BibitemShut {NoStop}%
	\bibitem [{\citenamefont {Fransson}\ \emph {et~al.}(2016)\citenamefont
		{Fransson}, \citenamefont {{Black-Schaffer}},\ and\ \citenamefont
		{Balatsky}}]{franssonMagnonDiracMaterials2016}%
	\BibitemOpen
	\bibfield  {author} {\bibinfo {author} {\bibfnamefont {J.}~\bibnamefont
			{Fransson}}, \bibinfo {author} {\bibfnamefont {A.~M.}\ \bibnamefont
			{{Black-Schaffer}}},\ and\ \bibinfo {author} {\bibfnamefont {A.~V.}\
			\bibnamefont {Balatsky}},\ }\bibfield  {title} {\bibinfo {title} {Magnon
			{{Dirac}} materials},\ }\href {https://doi.org/10.1103/PhysRevB.94.075401}
	{\bibfield  {journal} {\bibinfo  {journal} {Physical Review B}\ }\textbf
		{\bibinfo {volume} {94}},\ \bibinfo {pages} {075401} (\bibinfo {year}
		{2016})}\BibitemShut {NoStop}%
	\bibitem [{\citenamefont {Pershoguba}\ \emph {et~al.}(2018)\citenamefont
		{Pershoguba}, \citenamefont {Banerjee}, \citenamefont {Lashley},
		\citenamefont {Park}, \citenamefont {{\AA}gren}, \citenamefont {Aeppli},\
		and\ \citenamefont {Balatsky}}]{pershogubaDiracMagnonsHoneycomb2018}%
	\BibitemOpen
	\bibfield  {author} {\bibinfo {author} {\bibfnamefont {S.~S.}\ \bibnamefont
			{Pershoguba}}, \bibinfo {author} {\bibfnamefont {S.}~\bibnamefont
			{Banerjee}}, \bibinfo {author} {\bibfnamefont {J.~C.}\ \bibnamefont
			{Lashley}}, \bibinfo {author} {\bibfnamefont {J.}~\bibnamefont {Park}},
		\bibinfo {author} {\bibfnamefont {H.}~\bibnamefont {{\AA}gren}}, \bibinfo
		{author} {\bibfnamefont {G.}~\bibnamefont {Aeppli}},\ and\ \bibinfo {author}
		{\bibfnamefont {A.~V.}\ \bibnamefont {Balatsky}},\ }\bibfield  {title}
	{\bibinfo {title} {Dirac {{Magnons}} in {{Honeycomb Ferromagnets}}},\ }\href
	{https://doi.org/10.1103/PhysRevX.8.011010} {\bibfield  {journal} {\bibinfo
			{journal} {Physical Review X}\ }\textbf {\bibinfo {volume} {8}},\ \bibinfo
		{pages} {011010} (\bibinfo {year} {2018})}\BibitemShut {NoStop}%
	\bibitem [{\citenamefont {Li}\ \emph {et~al.}(2016)\citenamefont {Li},
		\citenamefont {Li}, \citenamefont {Kim}, \citenamefont {Balents},
		\citenamefont {Yu},\ and\ \citenamefont {Chen}}]{liWeylMagnonsBreathing2016}%
	\BibitemOpen
	\bibfield  {author} {\bibinfo {author} {\bibfnamefont {F.-Y.}\ \bibnamefont
			{Li}}, \bibinfo {author} {\bibfnamefont {Y.-D.}\ \bibnamefont {Li}}, \bibinfo
		{author} {\bibfnamefont {Y.~B.}\ \bibnamefont {Kim}}, \bibinfo {author}
		{\bibfnamefont {L.}~\bibnamefont {Balents}}, \bibinfo {author} {\bibfnamefont
			{Y.}~\bibnamefont {Yu}},\ and\ \bibinfo {author} {\bibfnamefont
			{G.}~\bibnamefont {Chen}},\ }\bibfield  {title} {\bibinfo {title} {Weyl
			magnons in breathing pyrochlore antiferromagnets},\ }\href
	{https://doi.org/10.1038/ncomms12691} {\bibfield  {journal} {\bibinfo
			{journal} {Nature Communications}\ }\textbf {\bibinfo {volume} {7}},\
		\bibinfo {pages} {12691} (\bibinfo {year} {2016})}\BibitemShut {NoStop}%
	\bibitem [{\citenamefont {Mook}\ \emph {et~al.}(2016)\citenamefont {Mook},
		\citenamefont {Henk},\ and\ \citenamefont
		{Mertig}}]{mookTunableMagnonWeyl2016}%
	\BibitemOpen
	\bibfield  {author} {\bibinfo {author} {\bibfnamefont {A.}~\bibnamefont
			{Mook}}, \bibinfo {author} {\bibfnamefont {J.}~\bibnamefont {Henk}},\ and\
		\bibinfo {author} {\bibfnamefont {I.}~\bibnamefont {Mertig}},\ }\bibfield
	{title} {\bibinfo {title} {Tunable {{Magnon Weyl Points}} in {{Ferromagnetic
					Pyrochlores}}},\ }\href {https://doi.org/10.1103/PhysRevLett.117.157204}
	{\bibfield  {journal} {\bibinfo  {journal} {Physical Review Letters}\
		}\textbf {\bibinfo {volume} {117}},\ \bibinfo {pages} {157204} (\bibinfo
		{year} {2016})}\BibitemShut {NoStop}%
	\bibitem [{\citenamefont {Li}\ \emph {et~al.}(2019)\citenamefont {Li},
		\citenamefont {Cao}, \citenamefont {Yan},\ and\ \citenamefont
		{Wang}}]{liHigherorderTopologicalSolitonic2019}%
	\BibitemOpen
	\bibfield  {author} {\bibinfo {author} {\bibfnamefont {Z.}~\bibnamefont
			{Li}}, \bibinfo {author} {\bibfnamefont {Y.}~\bibnamefont {Cao}}, \bibinfo
		{author} {\bibfnamefont {P.}~\bibnamefont {Yan}},\ and\ \bibinfo {author}
		{\bibfnamefont {X.}~\bibnamefont {Wang}},\ }\bibfield  {title} {\bibinfo
		{title} {Higher-order topological solitonic insulators},\ }\href
	{https://doi.org/10.1038/s41524-019-0246-4} {\bibfield  {journal} {\bibinfo
			{journal} {npj Computational Materials}\ }\textbf {\bibinfo {volume} {5}},\
		\bibinfo {pages} {1} (\bibinfo {year} {2019})}\BibitemShut {NoStop}%
	\bibitem [{\citenamefont {Hirosawa}\ \emph {et~al.}(2020)\citenamefont
		{Hirosawa}, \citenamefont {D{\'i}az}, \citenamefont {Klinovaja},\ and\
		\citenamefont {Loss}}]{hirosawaMagnonicQuadrupoleTopological2020}%
	\BibitemOpen
	\bibfield  {author} {\bibinfo {author} {\bibfnamefont {T.}~\bibnamefont
			{Hirosawa}}, \bibinfo {author} {\bibfnamefont {S.~A.}\ \bibnamefont
			{D{\'i}az}}, \bibinfo {author} {\bibfnamefont {J.}~\bibnamefont
			{Klinovaja}},\ and\ \bibinfo {author} {\bibfnamefont {D.}~\bibnamefont
			{Loss}},\ }\bibfield  {title} {\bibinfo {title} {Magnonic {{Quadrupole
					Topological Insulator}} in {{Antiskyrmion Crystals}}},\ }\href
	{https://doi.org/10.1103/PhysRevLett.125.207204} {\bibfield  {journal}
		{\bibinfo  {journal} {Physical Review Letters}\ }\textbf {\bibinfo {volume}
			{125}},\ \bibinfo {pages} {207204} (\bibinfo {year} {2020})}\BibitemShut
	{NoStop}%
	\bibitem [{\citenamefont {Mook}\ \emph
		{et~al.}(2021{\natexlab{b}})\citenamefont {Mook}, \citenamefont {D{\'i}az},
		\citenamefont {Klinovaja},\ and\ \citenamefont
		{Loss}}]{mookChiralHingeMagnons2021}%
	\BibitemOpen
	\bibfield  {author} {\bibinfo {author} {\bibfnamefont {A.}~\bibnamefont
			{Mook}}, \bibinfo {author} {\bibfnamefont {S.~A.}\ \bibnamefont {D{\'i}az}},
		\bibinfo {author} {\bibfnamefont {J.}~\bibnamefont {Klinovaja}},\ and\
		\bibinfo {author} {\bibfnamefont {D.}~\bibnamefont {Loss}},\ }\bibfield
	{title} {\bibinfo {title} {Chiral hinge magnons in second-order topological
			magnon insulators},\ }\href {https://doi.org/10.1103/PhysRevB.104.024406}
	{\bibfield  {journal} {\bibinfo  {journal} {Physical Review B}\ }\textbf
		{\bibinfo {volume} {104}},\ \bibinfo {pages} {024406} (\bibinfo {year}
		{2021}{\natexlab{b}})}\BibitemShut {NoStop}%
	\bibitem [{\citenamefont {Wang}\ \emph {et~al.}(2018)\citenamefont {Wang},
		\citenamefont {Zhang},\ and\ \citenamefont
		{Wang}}]{wangTopologicalMagnonicsParadigm2018}%
	\BibitemOpen
	\bibfield  {author} {\bibinfo {author} {\bibfnamefont {X.~S.}\ \bibnamefont
			{Wang}}, \bibinfo {author} {\bibfnamefont {H.~W.}\ \bibnamefont {Zhang}},\
		and\ \bibinfo {author} {\bibfnamefont {X.~R.}\ \bibnamefont {Wang}},\
	}\bibfield  {title} {\bibinfo {title} {Topological {{Magnonics}}: {{A
					Paradigm}} for {{Spin-Wave Manipulation}} and {{Device Design}}},\ }\href
	{https://doi.org/10.1103/PhysRevApplied.9.024029} {\bibfield  {journal}
		{\bibinfo  {journal} {Physical Review Applied}\ }\textbf {\bibinfo {volume}
			{9}},\ \bibinfo {pages} {024029} (\bibinfo {year} {2018})}\BibitemShut
	{NoStop}%
	\bibitem [{\citenamefont {R{\"u}ckriegel}\ \emph {et~al.}(2018)\citenamefont
		{R{\"u}ckriegel}, \citenamefont {Brataas},\ and\ \citenamefont
		{Duine}}]{ruckriegelBulkEdgeSpin2018}%
	\BibitemOpen
	\bibfield  {author} {\bibinfo {author} {\bibfnamefont {A.}~\bibnamefont
			{R{\"u}ckriegel}}, \bibinfo {author} {\bibfnamefont {A.}~\bibnamefont
			{Brataas}},\ and\ \bibinfo {author} {\bibfnamefont {R.~A.}\ \bibnamefont
			{Duine}},\ }\bibfield  {title} {\bibinfo {title} {Bulk and edge spin
			transport in topological magnon insulators},\ }\href
	{https://doi.org/10.1103/PhysRevB.97.081106} {\bibfield  {journal} {\bibinfo
			{journal} {Physical Review B}\ }\textbf {\bibinfo {volume} {97}},\ \bibinfo
		{pages} {081106(R)} (\bibinfo {year} {2018})}\BibitemShut {NoStop}%
	\bibitem [{\citenamefont {Wang}\ \emph {et~al.}(2020)\citenamefont {Wang},
		\citenamefont {Brataas},\ and\ \citenamefont
		{Troncoso}}]{wangBosonicBottIndex2020}%
	\BibitemOpen
	\bibfield  {author} {\bibinfo {author} {\bibfnamefont {X.~S.}\ \bibnamefont
			{Wang}}, \bibinfo {author} {\bibfnamefont {A.}~\bibnamefont {Brataas}},\ and\
		\bibinfo {author} {\bibfnamefont {R.~E.}\ \bibnamefont {Troncoso}},\
	}\bibfield  {title} {\bibinfo {title} {Bosonic {{Bott Index}} and
			{{Disorder-Induced Topological Transitions}} of {{Magnons}}},\ }\href
	{https://doi.org/10.1103/PhysRevLett.125.217202} {\bibfield  {journal}
		{\bibinfo  {journal} {Physical Review Letters}\ }\textbf {\bibinfo {volume}
			{125}},\ \bibinfo {pages} {217202} (\bibinfo {year} {2020})}\BibitemShut
	{NoStop}%
	\bibitem [{\citenamefont {Chisnell}\ \emph {et~al.}(2015)\citenamefont
		{Chisnell}, \citenamefont {Helton}, \citenamefont {Freedman}, \citenamefont
		{Singh}, \citenamefont {Bewley}, \citenamefont {Nocera},\ and\ \citenamefont
		{Lee}}]{chisnellTopologicalMagnonBands2015}%
	\BibitemOpen
	\bibfield  {author} {\bibinfo {author} {\bibfnamefont {R.}~\bibnamefont
			{Chisnell}}, \bibinfo {author} {\bibfnamefont {J.~S.}\ \bibnamefont
			{Helton}}, \bibinfo {author} {\bibfnamefont {D.~E.}\ \bibnamefont
			{Freedman}}, \bibinfo {author} {\bibfnamefont {D.~K.}\ \bibnamefont {Singh}},
		\bibinfo {author} {\bibfnamefont {R.~I.}\ \bibnamefont {Bewley}}, \bibinfo
		{author} {\bibfnamefont {D.~G.}\ \bibnamefont {Nocera}},\ and\ \bibinfo
		{author} {\bibfnamefont {Y.~S.}\ \bibnamefont {Lee}},\ }\bibfield  {title}
	{\bibinfo {title} {Topological {{Magnon Bands}} in a {{Kagome Lattice
					Ferromagnet}}},\ }\href {https://doi.org/10.1103/PhysRevLett.115.147201}
	{\bibfield  {journal} {\bibinfo  {journal} {Physical Review Letters}\
		}\textbf {\bibinfo {volume} {115}},\ \bibinfo {pages} {147201} (\bibinfo
		{year} {2015})}\BibitemShut {NoStop}%
	\bibitem [{\citenamefont {Chen}\ \emph {et~al.}(2018)\citenamefont {Chen},
		\citenamefont {Chung}, \citenamefont {Gao}, \citenamefont {Chen},
		\citenamefont {Stone}, \citenamefont {Kolesnikov}, \citenamefont {Huang},\
		and\ \citenamefont {Dai}}]{chenTopologicalSpinExcitations2018}%
	\BibitemOpen
	\bibfield  {author} {\bibinfo {author} {\bibfnamefont {L.}~\bibnamefont
			{Chen}}, \bibinfo {author} {\bibfnamefont {J.-H.}\ \bibnamefont {Chung}},
		\bibinfo {author} {\bibfnamefont {B.}~\bibnamefont {Gao}}, \bibinfo {author}
		{\bibfnamefont {T.}~\bibnamefont {Chen}}, \bibinfo {author} {\bibfnamefont
			{M.~B.}\ \bibnamefont {Stone}}, \bibinfo {author} {\bibfnamefont {A.~I.}\
			\bibnamefont {Kolesnikov}}, \bibinfo {author} {\bibfnamefont
			{Q.}~\bibnamefont {Huang}},\ and\ \bibinfo {author} {\bibfnamefont
			{P.}~\bibnamefont {Dai}},\ }\bibfield  {title} {\bibinfo {title} {Topological
			{{Spin Excitations}} in {{Honeycomb Ferromagnet}} {{${\mathrm{CrI}}_{3}$}}},\
	}\href {https://doi.org/10.1103/PhysRevX.8.041028} {\bibfield  {journal}
		{\bibinfo  {journal} {Physical Review X}\ }\textbf {\bibinfo {volume} {8}},\
		\bibinfo {pages} {041028} (\bibinfo {year} {2018})}\BibitemShut {NoStop}%
	\bibitem [{\citenamefont {Zhu}\ \emph {et~al.}(2021)\citenamefont {Zhu},
		\citenamefont {Zhang}, \citenamefont {Wang}, \citenamefont {{dos Santos}},
		\citenamefont {Song}, \citenamefont {Mueller}, \citenamefont {Schmalzl},
		\citenamefont {Schmidt}, \citenamefont {Ivanov}, \citenamefont {Park},
		\citenamefont {Xu}, \citenamefont {Ma}, \citenamefont {Lounis}, \citenamefont
		{Bl{\"u}gel}, \citenamefont {Mokrousov}, \citenamefont {Su},\ and\
		\citenamefont {Br{\"u}ckel}}]{zhuTopologicalMagnonInsulators2021}%
	\BibitemOpen
	\bibfield  {author} {\bibinfo {author} {\bibfnamefont {F.}~\bibnamefont
			{Zhu}}, \bibinfo {author} {\bibfnamefont {L.}~\bibnamefont {Zhang}}, \bibinfo
		{author} {\bibfnamefont {X.}~\bibnamefont {Wang}}, \bibinfo {author}
		{\bibfnamefont {F.~J.}\ \bibnamefont {{dos Santos}}}, \bibinfo {author}
		{\bibfnamefont {J.}~\bibnamefont {Song}}, \bibinfo {author} {\bibfnamefont
			{T.}~\bibnamefont {Mueller}}, \bibinfo {author} {\bibfnamefont
			{K.}~\bibnamefont {Schmalzl}}, \bibinfo {author} {\bibfnamefont {W.~F.}\
			\bibnamefont {Schmidt}}, \bibinfo {author} {\bibfnamefont {A.}~\bibnamefont
			{Ivanov}}, \bibinfo {author} {\bibfnamefont {J.~T.}\ \bibnamefont {Park}},
		\bibinfo {author} {\bibfnamefont {J.}~\bibnamefont {Xu}}, \bibinfo {author}
		{\bibfnamefont {J.}~\bibnamefont {Ma}}, \bibinfo {author} {\bibfnamefont
			{S.}~\bibnamefont {Lounis}}, \bibinfo {author} {\bibfnamefont
			{S.}~\bibnamefont {Bl{\"u}gel}}, \bibinfo {author} {\bibfnamefont
			{Y.}~\bibnamefont {Mokrousov}}, \bibinfo {author} {\bibfnamefont
			{Y.}~\bibnamefont {Su}},\ and\ \bibinfo {author} {\bibfnamefont
			{T.}~\bibnamefont {Br{\"u}ckel}},\ }\bibfield  {title} {\bibinfo {title}
		{Topological magnon insulators in two-dimensional van der {{Waals}}
			ferromagnets {{${\mathrm{CrSiTe}_{3}}$}} and {{${\mathrm{CrGeTe}_{3}}$}}:
			{{Toward}} intrinsic gap-tunability},\ }\href
	{https://doi.org/10.1126/sciadv.abi7532} {\bibfield  {journal} {\bibinfo
			{journal} {Science Advances}\ }\textbf {\bibinfo {volume} {7}},\ \bibinfo
		{pages} {eabi7532} (\bibinfo {year} {2021})}\BibitemShut {NoStop}%
	\bibitem [{\citenamefont {Weber}\ \emph {et~al.}(2022)\citenamefont {Weber},
		\citenamefont {Fobes}, \citenamefont {Waizner}, \citenamefont {Steffens},
		\citenamefont {Tucker}, \citenamefont {B{\"o}hm}, \citenamefont {Beddrich},
		\citenamefont {Franz}, \citenamefont {Gabold}, \citenamefont {Bewley},
		\citenamefont {Voneshen}, \citenamefont {Skoulatos}, \citenamefont {Georgii},
		\citenamefont {Ehlers}, \citenamefont {Bauer}, \citenamefont {Pfleiderer},
		\citenamefont {B{\"o}ni}, \citenamefont {Janoschek},\ and\ \citenamefont
		{Garst}}]{weberTopologicalMagnonBand2022}%
	\BibitemOpen
	\bibfield  {author} {\bibinfo {author} {\bibfnamefont {T.}~\bibnamefont
			{Weber}}, \bibinfo {author} {\bibfnamefont {D.~M.}\ \bibnamefont {Fobes}},
		\bibinfo {author} {\bibfnamefont {J.}~\bibnamefont {Waizner}}, \bibinfo
		{author} {\bibfnamefont {P.}~\bibnamefont {Steffens}}, \bibinfo {author}
		{\bibfnamefont {G.~S.}\ \bibnamefont {Tucker}}, \bibinfo {author}
		{\bibfnamefont {M.}~\bibnamefont {B{\"o}hm}}, \bibinfo {author}
		{\bibfnamefont {L.}~\bibnamefont {Beddrich}}, \bibinfo {author}
		{\bibfnamefont {C.}~\bibnamefont {Franz}}, \bibinfo {author} {\bibfnamefont
			{H.}~\bibnamefont {Gabold}}, \bibinfo {author} {\bibfnamefont
			{R.}~\bibnamefont {Bewley}}, \bibinfo {author} {\bibfnamefont
			{D.}~\bibnamefont {Voneshen}}, \bibinfo {author} {\bibfnamefont
			{M.}~\bibnamefont {Skoulatos}}, \bibinfo {author} {\bibfnamefont
			{R.}~\bibnamefont {Georgii}}, \bibinfo {author} {\bibfnamefont
			{G.}~\bibnamefont {Ehlers}}, \bibinfo {author} {\bibfnamefont
			{A.}~\bibnamefont {Bauer}}, \bibinfo {author} {\bibfnamefont
			{C.}~\bibnamefont {Pfleiderer}}, \bibinfo {author} {\bibfnamefont
			{P.}~\bibnamefont {B{\"o}ni}}, \bibinfo {author} {\bibfnamefont
			{M.}~\bibnamefont {Janoschek}},\ and\ \bibinfo {author} {\bibfnamefont
			{M.}~\bibnamefont {Garst}},\ }\bibfield  {title} {\bibinfo {title}
		{Topological magnon band structure of emergent {{Landau}} levels in a
			skyrmion lattice},\ }\href {https://doi.org/10.1126/science.abe4441}
	{\bibfield  {journal} {\bibinfo  {journal} {Science}\ }\textbf {\bibinfo
			{volume} {375}},\ \bibinfo {pages} {1025} (\bibinfo {year}
		{2022})}\BibitemShut {NoStop}%
	\bibitem [{\citenamefont {Vi{\~n}as~Bostr{\"o}m}\ \emph
		{et~al.}(2023)\citenamefont {Vi{\~n}as~Bostr{\"o}m}, \citenamefont {Parvini},
		\citenamefont {McIver}, \citenamefont {Rubio}, \citenamefont {Kusminskiy},\
		and\ \citenamefont {Sentef}}]{vinasbostromDirectOpticalProbe2023}%
	\BibitemOpen
	\bibfield  {author} {\bibinfo {author} {\bibfnamefont {E.}~\bibnamefont
			{Vi{\~n}as~Bostr{\"o}m}}, \bibinfo {author} {\bibfnamefont {T.~S.}\
			\bibnamefont {Parvini}}, \bibinfo {author} {\bibfnamefont {J.~W.}\
			\bibnamefont {McIver}}, \bibinfo {author} {\bibfnamefont {A.}~\bibnamefont
			{Rubio}}, \bibinfo {author} {\bibfnamefont {S.~V.}\ \bibnamefont
			{Kusminskiy}},\ and\ \bibinfo {author} {\bibfnamefont {M.~A.}\ \bibnamefont
			{Sentef}},\ }\bibfield  {title} {\bibinfo {title} {Direct {{Optical Probe}}
			of {{Magnon Topology}} in {{Two-Dimensional Quantum Magnets}}},\ }\href
	{https://doi.org/10.1103/PhysRevLett.130.026701} {\bibfield  {journal}
		{\bibinfo  {journal} {Physical Review Letters}\ }\textbf {\bibinfo {volume}
			{130}},\ \bibinfo {pages} {026701} (\bibinfo {year} {2023})}\BibitemShut
	{NoStop}%
	\bibitem [{\citenamefont {Lu}\ and\ \citenamefont
		{Lu}(2018)}]{luMagnonBandTopology2018}%
	\BibitemOpen
	\bibfield  {author} {\bibinfo {author} {\bibfnamefont {F.}~\bibnamefont
			{Lu}}\ and\ \bibinfo {author} {\bibfnamefont {Y.-M.}\ \bibnamefont {Lu}},\
	}\href {https://doi.org/10.48550/arXiv.1807.05232} {\bibinfo {title} {Magnon
			band topology in spin-orbital coupled magnets: Classification and application
			to $\alpha$-rucl$_3$}} (\bibinfo {year} {2018}),\ \Eprint
	{https://arxiv.org/abs/1807.05232} {arxiv:1807.05232} \BibitemShut {NoStop}%
	\bibitem [{\citenamefont {Xu}\ \emph {et~al.}(2020)\citenamefont {Xu},
		\citenamefont {Flynn}, \citenamefont {Alase}, \citenamefont {Cobanera},
		\citenamefont {Viola},\ and\ \citenamefont
		{Ortiz}}]{xuSquaringFermionThreefold2020}%
	\BibitemOpen
	\bibfield  {author} {\bibinfo {author} {\bibfnamefont {Q.-R.}\ \bibnamefont
			{Xu}}, \bibinfo {author} {\bibfnamefont {V.~P.}\ \bibnamefont {Flynn}},
		\bibinfo {author} {\bibfnamefont {A.}~\bibnamefont {Alase}}, \bibinfo
		{author} {\bibfnamefont {E.}~\bibnamefont {Cobanera}}, \bibinfo {author}
		{\bibfnamefont {L.}~\bibnamefont {Viola}},\ and\ \bibinfo {author}
		{\bibfnamefont {G.}~\bibnamefont {Ortiz}},\ }\bibfield  {title} {\bibinfo
		{title} {Squaring the fermion: {{The}} threefold way and the fate of zero
			modes},\ }\href {https://doi.org/10.1103/PhysRevB.102.125127} {\bibfield
		{journal} {\bibinfo  {journal} {Physical Review B}\ }\textbf {\bibinfo
			{volume} {102}},\ \bibinfo {pages} {125127} (\bibinfo {year}
		{2020})}\BibitemShut {NoStop}%
	\bibitem [{\citenamefont {Mena}\ \emph {et~al.}(2014)\citenamefont {Mena},
		\citenamefont {Perry}, \citenamefont {Perring}, \citenamefont {Le},
		\citenamefont {Guerrero}, \citenamefont {Storni}, \citenamefont {Adroja},
		\citenamefont {R{\"u}egg},\ and\ \citenamefont
		{McMorrow}}]{menaSpinWaveSpectrumQuantum2014}%
	\BibitemOpen
	\bibfield  {author} {\bibinfo {author} {\bibfnamefont {M.}~\bibnamefont
			{Mena}}, \bibinfo {author} {\bibfnamefont {R.~S.}\ \bibnamefont {Perry}},
		\bibinfo {author} {\bibfnamefont {T.~G.}\ \bibnamefont {Perring}}, \bibinfo
		{author} {\bibfnamefont {M.~D.}\ \bibnamefont {Le}}, \bibinfo {author}
		{\bibfnamefont {S.}~\bibnamefont {Guerrero}}, \bibinfo {author}
		{\bibfnamefont {M.}~\bibnamefont {Storni}}, \bibinfo {author} {\bibfnamefont
			{D.~T.}\ \bibnamefont {Adroja}}, \bibinfo {author} {\bibfnamefont {{\relax
					Ch}.}~\bibnamefont {R{\"u}egg}},\ and\ \bibinfo {author} {\bibfnamefont
			{D.~F.}\ \bibnamefont {McMorrow}},\ }\bibfield  {title} {\bibinfo {title}
		{Spin-{{Wave Spectrum}} of the {{Quantum Ferromagnet}} on the {{Pyrochlore
					Lattice}} {{${\mathrm{Lu}}_{2}{\mathrm{V}}_{2}{\mathrm{O}}_{7}$}}},\ }\href
	{https://doi.org/10.1103/PhysRevLett.113.047202} {\bibfield  {journal}
		{\bibinfo  {journal} {Physical Review Letters}\ }\textbf {\bibinfo {volume}
			{113}},\ \bibinfo {pages} {047202} (\bibinfo {year} {2014})}\BibitemShut
	{NoStop}%
	\bibitem [{Note1()}]{Note1}%
	\BibitemOpen
	\bibinfo {note} {Because of the specific geometry considered here, where the
		magnetization is perpendicular to the plane, one would need to make use of
		the anomalous spin Hall effect in a ferromagnetic heavy metal, such as
		permalloy \cite {dasSpinInjectionDetection2017}.}\BibitemShut {Stop}%
	\bibitem [{Note2()}]{Note2}%
	\BibitemOpen
	\bibinfo {note} {See Supplemental Material for the details of the Haldane model and the effects of
		anisotropy, a full discussion on the particle-hole symmetry, details on the
		LLG simulations and transmission calculations, additional sources of disorder
		and an estimation of the energy scales.}\BibitemShut
	{Stop}%
	\bibitem [{\citenamefont {Harms}\ \emph {et~al.}(2022)\citenamefont {Harms},
		\citenamefont {Yuan},\ and\ \citenamefont {Duine}}]{harmsAntimagnonics2022}%
	\BibitemOpen
	\bibfield  {author} {\bibinfo {author} {\bibfnamefont {J.~S.}\ \bibnamefont
			{Harms}}, \bibinfo {author} {\bibfnamefont {H.~Y.}\ \bibnamefont {Yuan}},\
		and\ \bibinfo {author} {\bibfnamefont {R.~A.}\ \bibnamefont {Duine}},\ }\href
	{https://doi.org/10.48550/arXiv.2210.16698} {\bibinfo {title}
		{Antimagnonics}} (\bibinfo {year} {2022}),\ \Eprint
	{https://arxiv.org/abs/2210.16698} {arxiv:2210.16698} \BibitemShut {NoStop}%
	\bibitem [{\citenamefont {Lein}\ and\ \citenamefont
		{Sato}(2019)}]{leinKreinSchrodingerFormalismBosonic2019}%
	\BibitemOpen
	\bibfield  {author} {\bibinfo {author} {\bibfnamefont {M.}~\bibnamefont
			{Lein}}\ and\ \bibinfo {author} {\bibfnamefont {K.}~\bibnamefont {Sato}},\
	}\bibfield  {title} {\bibinfo {title} {Krein-{{Schr\"odinger}} formalism of
			bosonic {{Bogoliubov--de Gennes}} and certain classical systems and their
			topological classification},\ }\href
	{https://doi.org/10.1103/PhysRevB.100.075414} {\bibfield  {journal} {\bibinfo
			{journal} {Physical Review B}\ }\textbf {\bibinfo {volume} {100}},\ \bibinfo
		{pages} {075414} (\bibinfo {year} {2019})}\BibitemShut {NoStop}%
	\bibitem [{\citenamefont {Gunnink}\ \emph {et~al.}(2021)\citenamefont
		{Gunnink}, \citenamefont {Duine},\ and\ \citenamefont
		{R{\"u}ckriegel}}]{gunninkTheoryElectricalDetection2021}%
	\BibitemOpen
	\bibfield  {author} {\bibinfo {author} {\bibfnamefont {P.~M.}\ \bibnamefont
			{Gunnink}}, \bibinfo {author} {\bibfnamefont {R.~A.}\ \bibnamefont {Duine}},\
		and\ \bibinfo {author} {\bibfnamefont {A.}~\bibnamefont {R{\"u}ckriegel}},\
	}\bibfield  {title} {\bibinfo {title} {Theory for electrical detection of the
			magnon {{Hall}} effect induced by dipolar interactions},\ }\href
	{https://doi.org/10.1103/PhysRevB.103.214426} {\bibfield  {journal} {\bibinfo
			{journal} {Physical Review B}\ }\textbf {\bibinfo {volume} {103}},\ \bibinfo
		{pages} {214426} (\bibinfo {year} {2021})}\BibitemShut {NoStop}%
	\bibitem [{\citenamefont {Chumak}\ \emph {et~al.}(2012)\citenamefont {Chumak},
		\citenamefont {Serga}, \citenamefont {Jungfleisch}, \citenamefont {Neb},
		\citenamefont {Bozhko}, \citenamefont {Tiberkevich},\ and\ \citenamefont
		{Hillebrands}}]{chumakDirectDetectionMagnon2012}%
	\BibitemOpen
	\bibfield  {author} {\bibinfo {author} {\bibfnamefont {A.~V.}\ \bibnamefont
			{Chumak}}, \bibinfo {author} {\bibfnamefont {A.~A.}\ \bibnamefont {Serga}},
		\bibinfo {author} {\bibfnamefont {M.~B.}\ \bibnamefont {Jungfleisch}},
		\bibinfo {author} {\bibfnamefont {R.}~\bibnamefont {Neb}}, \bibinfo {author}
		{\bibfnamefont {D.~A.}\ \bibnamefont {Bozhko}}, \bibinfo {author}
		{\bibfnamefont {V.~S.}\ \bibnamefont {Tiberkevich}},\ and\ \bibinfo {author}
		{\bibfnamefont {B.}~\bibnamefont {Hillebrands}},\ }\bibfield  {title}
	{\bibinfo {title} {Direct detection of magnon spin transport by the inverse
			spin {{Hall}} effect},\ }\href {https://doi.org/10.1063/1.3689787} {\bibfield
		{journal} {\bibinfo  {journal} {Applied Physics Letters}\ }\textbf {\bibinfo
			{volume} {100}},\ \bibinfo {pages} {082405} (\bibinfo {year}
		{2012})}\BibitemShut {NoStop}%
	\bibitem [{\citenamefont {Barker}\ and\ \citenamefont
		{Bauer}(2016)}]{barkerThermalSpinDynamics2016}%
	\BibitemOpen
	\bibfield  {author} {\bibinfo {author} {\bibfnamefont {J.}~\bibnamefont
			{Barker}}\ and\ \bibinfo {author} {\bibfnamefont {G.~E.~W.}\ \bibnamefont
			{Bauer}},\ }\bibfield  {title} {\bibinfo {title} {Thermal {{Spin Dynamics}}
			of {{Yttrium Iron Garnet}}},\ }\href
	{https://doi.org/10.1103/PhysRevLett.117.217201} {\bibfield  {journal}
		{\bibinfo  {journal} {Physical Review Letters}\ }\textbf {\bibinfo {volume}
			{117}},\ \bibinfo {pages} {217201} (\bibinfo {year} {2016})}\BibitemShut
	{NoStop}%
	\bibitem [{\citenamefont {Chumak}\ \emph {et~al.}(2015)\citenamefont {Chumak},
		\citenamefont {Vasyuchka}, \citenamefont {Serga},\ and\ \citenamefont
		{Hillebrands}}]{chumakMagnonSpintronics2015}%
	\BibitemOpen
	\bibfield  {author} {\bibinfo {author} {\bibfnamefont {A.~V.}\ \bibnamefont
			{Chumak}}, \bibinfo {author} {\bibfnamefont {V.~I.}\ \bibnamefont
			{Vasyuchka}}, \bibinfo {author} {\bibfnamefont {A.~A.}\ \bibnamefont
			{Serga}},\ and\ \bibinfo {author} {\bibfnamefont {B.}~\bibnamefont
			{Hillebrands}},\ }\bibfield  {title} {\bibinfo {title} {Magnon spintronics},\
	}\href {https://doi.org/10.1038/nphys3347} {\bibfield  {journal} {\bibinfo
			{journal} {Nature Physics}\ }\textbf {\bibinfo {volume} {11}},\ \bibinfo
		{pages} {453} (\bibinfo {year} {2015})}\BibitemShut {NoStop}%
	\bibitem [{\citenamefont {Vlaminck}\ and\ \citenamefont
		{Bailleul}(2010)}]{vlaminckSpinwaveTransductionSubmicrometer2010}%
	\BibitemOpen
	\bibfield  {author} {\bibinfo {author} {\bibfnamefont {V.}~\bibnamefont
			{Vlaminck}}\ and\ \bibinfo {author} {\bibfnamefont {M.}~\bibnamefont
			{Bailleul}},\ }\bibfield  {title} {\bibinfo {title} {Spin-wave transduction
			at the submicrometer scale: {{Experiment}} and modeling},\ }\href
	{https://doi.org/10.1103/PhysRevB.81.014425} {\bibfield  {journal} {\bibinfo
			{journal} {Physical Review B}\ }\textbf {\bibinfo {volume} {81}},\ \bibinfo
		{pages} {014425} (\bibinfo {year} {2010})}\BibitemShut {NoStop}%
	\bibitem [{\citenamefont {Shindou}\ \emph
		{et~al.}(2013{\natexlab{b}})\citenamefont {Shindou}, \citenamefont {Ohe},
		\citenamefont {Matsumoto}, \citenamefont {Murakami},\ and\ \citenamefont
		{Saitoh}}]{shindouChiralSpinwaveEdge2013}%
	\BibitemOpen
	\bibfield  {author} {\bibinfo {author} {\bibfnamefont {R.}~\bibnamefont
			{Shindou}}, \bibinfo {author} {\bibfnamefont {J.-i.}\ \bibnamefont {Ohe}},
		\bibinfo {author} {\bibfnamefont {R.}~\bibnamefont {Matsumoto}}, \bibinfo
		{author} {\bibfnamefont {S.}~\bibnamefont {Murakami}},\ and\ \bibinfo
		{author} {\bibfnamefont {E.}~\bibnamefont {Saitoh}},\ }\bibfield  {title}
	{\bibinfo {title} {Chiral spin-wave edge modes in dipolar magnetic thin
			films},\ }\href {https://doi.org/10.1103/PhysRevB.87.174402} {\bibfield
		{journal} {\bibinfo  {journal} {Physical Review B}\ }\textbf {\bibinfo
			{volume} {87}},\ \bibinfo {pages} {174402} (\bibinfo {year}
		{2013}{\natexlab{b}})}\BibitemShut {NoStop}%
	\bibitem [{\citenamefont {Su}\ and\ \citenamefont
		{Wang}(2017)}]{suChiralAnomalyWeyl2017}%
	\BibitemOpen
	\bibfield  {author} {\bibinfo {author} {\bibfnamefont {Y.}~\bibnamefont
			{Su}}\ and\ \bibinfo {author} {\bibfnamefont {X.~R.}\ \bibnamefont {Wang}},\
	}\bibfield  {title} {\bibinfo {title} {Chiral anomaly of {{Weyl}} magnons in
			stacked honeycomb ferromagnets},\ }\href
	{https://doi.org/10.1103/PhysRevB.96.104437} {\bibfield  {journal} {\bibinfo
			{journal} {Physical Review B}\ }\textbf {\bibinfo {volume} {96}},\ \bibinfo
		{pages} {104437} (\bibinfo {year} {2017})}\BibitemShut {NoStop}%
	\bibitem [{\citenamefont {Liu}\ and\ \citenamefont
		{Shi}(2019)}]{liuMagnonQuantumAnomalies2019}%
	\BibitemOpen
	\bibfield  {author} {\bibinfo {author} {\bibfnamefont {T.}~\bibnamefont
			{Liu}}\ and\ \bibinfo {author} {\bibfnamefont {Z.}~\bibnamefont {Shi}},\
	}\bibfield  {title} {\bibinfo {title} {Magnon quantum anomalies in {{Weyl}}
			ferromagnets},\ }\href {https://doi.org/10.1103/PhysRevB.99.214413}
	{\bibfield  {journal} {\bibinfo  {journal} {Physical Review B}\ }\textbf
		{\bibinfo {volume} {99}},\ \bibinfo {pages} {214413} (\bibinfo {year}
		{2019})}\BibitemShut {NoStop}%
	\bibitem [{\citenamefont {Yang}\ \emph {et~al.}(2015)\citenamefont {Yang},
		\citenamefont {Gao}, \citenamefont {Shi}, \citenamefont {Lin}, \citenamefont
		{Gao}, \citenamefont {Chong},\ and\ \citenamefont
		{Zhang}}]{yangTopologicalAcoustics2015}%
	\BibitemOpen
	\bibfield  {author} {\bibinfo {author} {\bibfnamefont {Z.}~\bibnamefont
			{Yang}}, \bibinfo {author} {\bibfnamefont {F.}~\bibnamefont {Gao}}, \bibinfo
		{author} {\bibfnamefont {X.}~\bibnamefont {Shi}}, \bibinfo {author}
		{\bibfnamefont {X.}~\bibnamefont {Lin}}, \bibinfo {author} {\bibfnamefont
			{Z.}~\bibnamefont {Gao}}, \bibinfo {author} {\bibfnamefont {Y.}~\bibnamefont
			{Chong}},\ and\ \bibinfo {author} {\bibfnamefont {B.}~\bibnamefont {Zhang}},\
	}\bibfield  {title} {\bibinfo {title} {Topological {{Acoustics}}},\ }\href
	{https://doi.org/10.1103/PhysRevLett.114.114301} {\bibfield  {journal}
		{\bibinfo  {journal} {Physical Review Letters}\ }\textbf {\bibinfo {volume}
			{114}},\ \bibinfo {pages} {114301} (\bibinfo {year} {2015})}\BibitemShut
	{NoStop}%
	\bibitem [{\citenamefont {Wang}\ \emph {et~al.}(2015)\citenamefont {Wang},
		\citenamefont {Lu},\ and\ \citenamefont
		{Bertoldi}}]{wangTopologicalPhononicCrystals2015}%
	\BibitemOpen
	\bibfield  {author} {\bibinfo {author} {\bibfnamefont {P.}~\bibnamefont
			{Wang}}, \bibinfo {author} {\bibfnamefont {L.}~\bibnamefont {Lu}},\ and\
		\bibinfo {author} {\bibfnamefont {K.}~\bibnamefont {Bertoldi}},\ }\bibfield
	{title} {\bibinfo {title} {Topological {{Phononic Crystals}} with {{One-Way
					Elastic Edge Waves}}},\ }\href
	{https://doi.org/10.1103/PhysRevLett.115.104302} {\bibfield  {journal}
		{\bibinfo  {journal} {Physical Review Letters}\ }\textbf {\bibinfo {volume}
			{115}},\ \bibinfo {pages} {104302} (\bibinfo {year} {2015})}\BibitemShut
	{NoStop}%
	\bibitem [{\citenamefont {Onose}\ \emph {et~al.}(2010)\citenamefont {Onose},
		\citenamefont {Ideue}, \citenamefont {Katsura}, \citenamefont {Shiomi},
		\citenamefont {Nagaosa},\ and\ \citenamefont
		{Tokura}}]{onoseObservationMagnonHall2010}%
	\BibitemOpen
	\bibfield  {author} {\bibinfo {author} {\bibfnamefont {Y.}~\bibnamefont
			{Onose}}, \bibinfo {author} {\bibfnamefont {T.}~\bibnamefont {Ideue}},
		\bibinfo {author} {\bibfnamefont {H.}~\bibnamefont {Katsura}}, \bibinfo
		{author} {\bibfnamefont {Y.}~\bibnamefont {Shiomi}}, \bibinfo {author}
		{\bibfnamefont {N.}~\bibnamefont {Nagaosa}},\ and\ \bibinfo {author}
		{\bibfnamefont {Y.}~\bibnamefont {Tokura}},\ }\bibfield  {title} {\bibinfo
		{title} {Observation of the {{Magnon Hall Effect}}},\ }\href
	{https://doi.org/10.1126/science.1188260} {\bibfield  {journal} {\bibinfo
			{journal} {Science}\ }\textbf {\bibinfo {volume} {329}},\ \bibinfo {pages}
		{297} (\bibinfo {year} {2010})}\BibitemShut {NoStop}%
	\bibitem [{\citenamefont {Murakami}\ and\ \citenamefont
		{Okamoto}(2017)}]{murakamiThermalHallEffect2017}%
	\BibitemOpen
	\bibfield  {author} {\bibinfo {author} {\bibfnamefont {S.}~\bibnamefont
			{Murakami}}\ and\ \bibinfo {author} {\bibfnamefont {A.}~\bibnamefont
			{Okamoto}},\ }\bibfield  {title} {\bibinfo {title} {Thermal {{Hall Effect}}
			of {{Magnons}}},\ }\href {https://doi.org/10.7566/JPSJ.86.011010} {\bibfield
		{journal} {\bibinfo  {journal} {Journal of the Physical Society of Japan}\
		}\textbf {\bibinfo {volume} {86}},\ \bibinfo {pages} {011010} (\bibinfo
		{year} {2017})}\BibitemShut {NoStop}%
	\bibitem [{\citenamefont {Kovalev}\ and\ \citenamefont
		{Zyuzin}(2016)}]{kovalevSpinTorqueNernst2016}%
	\BibitemOpen
	\bibfield  {author} {\bibinfo {author} {\bibfnamefont {A.~A.}\ \bibnamefont
			{Kovalev}}\ and\ \bibinfo {author} {\bibfnamefont {V.}~\bibnamefont
			{Zyuzin}},\ }\bibfield  {title} {\bibinfo {title} {Spin torque and {{Nernst}}
			effects in {{Dzyaloshinskii-Moriya}} ferromagnets},\ }\href
	{https://doi.org/10.1103/PhysRevB.93.161106} {\bibfield  {journal} {\bibinfo
			{journal} {Physical Review B}\ }\textbf {\bibinfo {volume} {93}},\ \bibinfo
		{pages} {161106(R)} (\bibinfo {year} {2016})}\BibitemShut {NoStop}%
	\bibitem [{\citenamefont {Das}\ \emph {et~al.}(2017)\citenamefont {Das},
		\citenamefont {Schoemaker}, \citenamefont {{van Wees}},\ and\ \citenamefont
		{{Vera-Marun}}}]{dasSpinInjectionDetection2017}%
	\BibitemOpen
	\bibfield  {author} {\bibinfo {author} {\bibfnamefont {K.~S.}\ \bibnamefont
			{Das}}, \bibinfo {author} {\bibfnamefont {W.~Y.}\ \bibnamefont {Schoemaker}},
		\bibinfo {author} {\bibfnamefont {B.~J.}\ \bibnamefont {{van Wees}}},\ and\
		\bibinfo {author} {\bibfnamefont {I.~J.}\ \bibnamefont {{Vera-Marun}}},\
	}\bibfield  {title} {\bibinfo {title} {Spin injection and detection via the
			anomalous spin {{Hall}} effect of a ferromagnetic metal},\ }\href
	{https://doi.org/10.1103/PhysRevB.96.220408} {\bibfield  {journal} {\bibinfo
			{journal} {Physical Review B}\ }\textbf {\bibinfo {volume} {96}},\ \bibinfo
		{pages} {220408(R)} (\bibinfo {year} {2017})}\BibitemShut {NoStop}%
	\bibitem [{\citenamefont {Fukui}\ \emph {et~al.}(2005)\citenamefont {Fukui},
		\citenamefont {Hatsugai},\ and\ \citenamefont
		{Suzuki}}]{fukuiChernNumbersDiscretized2005}%
	\BibitemOpen
	\bibfield  {author} {\bibinfo {author} {\bibfnamefont {T.}~\bibnamefont
			{Fukui}}, \bibinfo {author} {\bibfnamefont {Y.}~\bibnamefont {Hatsugai}},\
		and\ \bibinfo {author} {\bibfnamefont {H.}~\bibnamefont {Suzuki}},\
	}\bibfield  {title} {\bibinfo {title} {Chern {{Numbers}} in {{Discretized
					Brillouin Zone}}: {{Efficient Method}} of {{Computing}} ({{Spin}}) {{Hall
					Conductances}}},\ }\href {https://doi.org/10.1143/JPSJ.74.1674} {\bibfield
		{journal} {\bibinfo  {journal} {Journal of the Physical Society of Japan}\
		}\textbf {\bibinfo {volume} {74}},\ \bibinfo {pages} {1674} (\bibinfo {year}
		{2005})}\BibitemShut {NoStop}%
	\bibitem [{\citenamefont {Castro}\ \emph {et~al.}(2015)\citenamefont {Castro},
		\citenamefont {{L{\'o}pez-Sancho}},\ and\ \citenamefont
		{Vozmediano}}]{castroAndersonLocalizationTopological2015}%
	\BibitemOpen
	\bibfield  {author} {\bibinfo {author} {\bibfnamefont {E.~V.}\ \bibnamefont
			{Castro}}, \bibinfo {author} {\bibfnamefont {M.~P.}\ \bibnamefont
			{{L{\'o}pez-Sancho}}},\ and\ \bibinfo {author} {\bibfnamefont {M.~A.~H.}\
			\bibnamefont {Vozmediano}},\ }\bibfield  {title} {\bibinfo {title} {Anderson
			localization and topological transition in {{Chern}} insulators},\ }\href
	{https://doi.org/10.1103/PhysRevB.92.085410} {\bibfield  {journal} {\bibinfo
			{journal} {Physical Review B}\ }\textbf {\bibinfo {volume} {92}},\ \bibinfo
		{pages} {085410} (\bibinfo {year} {2015})}\BibitemShut {NoStop}%
\end{thebibliography}
\end{document}